\documentclass[aps,prd,groupedaddress,preprint,eqsecnum,nofootinbib]{revtex4}
\usepackage{graphicx,epsf,amssymb,amsbsy,amsfonts,amssymb,amsmath,physics,courier,hyperref}

\def\be{\begin{equation}}
\def\ee{\end{equation}}
\def\bea{\begin{eqnarray}}
\def\eea{\end{eqnarray}}

\def\d{\delta}
\def\e{\epsilon}
\def\f{\phi}

\def\bg{\bar{g}}
\def\beq{\begin{eqnarray}}\def\eeq{\end{eqnarray}}
\def\ba#1\ea{\begin{align}#1\end{align}}
\def\bg#1\eg{\begin{gather}#1\end{gather}}
\def\bm#1\em{\begin{multline}#1\end{multline}}
\def\bmd#1\emd{\begin{multlined}#1\end{multlined}}

\def\d{\delta}
\def\D{\Delta}
\def\e{\epsilon}

\def\pa{\partial}

\def\({\left(}
\def\){\right)}
\def\[{\left[}
\def\]{\right]}

\def\d{\delta}

\def\no{\nonumber}
\def\f{\frac}

\def\D{\Delta}
\def\Om{\Omega}

\pdfoutput=1

\begin{document}
\hfuzz 9pt
\title{Conformal properties of soft-operators - 1 : Use of null-states }
\author{Shamik Banerjee, Pranjal Pandey, Partha Paul}
\affiliation{Institute of Physics, \\ Sachivalaya Marg, Bhubaneshwar, India-751005 \\ and \\ Homi Bhabha National Institute, Anushakti Nagar, Mumbai, India-400085}

\email{banerjeeshamik.phy@gmail.com, pranofmpvm@gmail.com, pl.partha13@gmail.com }

\begin{abstract} 
Soft-operators are (roughly speaking) zero energy massless particles which live on the celestial sphere in Minkowski space. The Lorentz group acts on the celestial sphere by conformal transformation and the soft-operators transform as \emph{conformal primary} operators of various dimension and spin. Working in space-time dimensions $D=4$ and $6$, in this paper, we study some properties of the conformal representations with (leading) soft photon and graviton as the highest weight operators. Typically these representations contain null-vectors. We argue, from the $S$--matrix point of view, that infinite dimensional asymptotic symmetries and conformal invariance require us to set these null-vectors to zero. As a result, the corresponding soft-operator satisfies linear PDE on the celestial sphere. Curiously, these PDEs are equations of motion of Euclidean gauge theories on the celestial sphere with \textit{scalar} gauge-invariance, i.e, the gauge parameter is a scalar field on the sphere. These are probably related to large $U(1)$ and supertranslation transformations at infinity. Now, the PDE satisfied by the soft-operator can be converted into PDE for the $S$--matrix elements with the insertion of the soft-operator. These equations can then  be solved subject to appropriate boundary conditions on the celestial sphere, provided by (Lorentz) conformal invariance. The resulting soft $S$--matrix elements have an interesting "pure-gauge" form and are determined in terms of a single scalar function. Heuristically speaking, the role of the null-state decoupling is to reduce the number of degrees of freedom or polarisation states of soft photon and graviton to \emph{one}, given effectively by a single scalar function. This reduction in the number of degrees of freedom makes the Ward-identity for the asymptotic symmetry almost integrable. The result of the integration, which we are not able to perform completely, should of course be Weinberg's soft-theorem. Finally, we comment on the resemblance of all of these things to quantization of fundamental strings.
\end{abstract}

\maketitle
\tableofcontents

\section{Introduction}
In asymptotically flat space-time a very interesting connection exists \cite{Strominger:2013lka, Strominger:2013jfa, He, Strominger:2014pwa, He:2014cra, Kapec:2015ena,Kapec:2014zla,Kapec:2015vwa,Pate:2017fgt, Campiglia:2015qka, Campiglia:2015kxa,Strominger:2017zoo} between soft-theorems and Ward identities of infinite dimensional (asymptotic) global symmetries. In even space-time dimensions conformal techniques are useful for studying this correspondence. In this subject, one basic fact is that the \emph{Lorentz group $SO(D-1,1)$ acts on the celestial sphere \footnote{Celestial sphere of a point in Minkowski space is essentially the space of null directions at that point. This space is topologically a sphere.} $S^{D-2}$ as the conformal group}. In fact, there are interesting operators in the theory, called soft operators \cite{Strominger:2013lka, Strominger:2013jfa, He, Strominger:2014pwa, He:2014cra, Kapec:2015ena,Kapec:2014zla,Kapec:2015vwa,Pate:2017fgt, Campiglia:2015qka, Campiglia:2015kxa,Kapec:2017gsg,Banerjee:2018fgd}, which live on the celestial sphere and transform as conformal primary under the (Lorentz) conformal group. It has also been conjectured \cite{Kapec:2016jld,Kapec:2017gsg,Pasterski:2016qvg,Pasterski:2017kqt,Pasterski:2017ylz,Cheung:2016iub,deBoer:2003vf,Banerjee:2018gce,Banerjee:2018fgd,Cardona:2017keg,Lam:2017ofc,Banerjee:2017jeg,Schreiber:2017jsr,Donnay:2018neh,Bagchi:2016bcd} that there is an Euclidean CFT on the celestial sphere which holographically computes the $S$--matrix in asymptotically flat space-time. Motivated by this conjecture, we try to understand in this paper the interplay between the conformal field theory structure and the infinite dimensional global symmetries. One way to get some hints of this interplay is to try to derive the soft-theorems from Ward-identities \emph{in a pure S--matrix theory}. By "pure $S$--matrix" we mean an asymptotic description. Therefore bulk equations of motion -- classical or quantum -- or the concept of bulk gauge symmetry are not naturally defined in this context. It seems that such an approach may not be completely unrealistic because we now have infinite dimensional global symmetries which act on the space of $S$--matrix elements. Moreover, these global symmetries are also asymptotic symmetries \cite{Strominger:2013lka,Strominger:2013jfa,Sachs:1962zza} and so, roughly speaking, constitute the subset of the bulk gauge transformations which act on the \emph{physical} Hilbert space. So at least in principle, nothing should be lost if we focus only on this physical part. Let us now briefly describe how we approach this problem. 

Instead of deriving the Ward-identities in a specific bulk theory, in this paper we take them as \emph{postulates}. For our purpose, it is useful to think of a theory as a collection of $S$--matrix elements -- the natural observables in a holographic description of asymptotically flat space-time. The \emph{global} symmetries act on this collection. Since we are talking about $S$--matrix elements, it is convenient to introduce physical asymptotic states. These asymptotic states determine the operator-content of the theory. In the best possible situation both the operator content and the $S$--matrix elements will be determined to a large extent by the symmetries themselves. For example, the infinite dimensional asymptotic $U(1)$ symmetry \cite{Strominger:2013lka, He:2014cra, Kapec:2015ena,Kapec:2014zla,Campiglia:2015qka, Strominger:2017zoo} is found in theories which have massless spin--$1$ particle or photon. So if we do not have the (soft) photon operator in the theory then the theory cannot have the $U(1)$ symmetry. Therefore, it is not too unreasonable to expect that in a holographic -- dual description the infinite dimensional \emph{global} $U(1)$ symmetry will predict the existence of (soft) photons. 

Now keeping these things in mind we introduce, besides the creation-annihilation operators for finite energy (physical) excitations, an infinite number of conformal primary operators of arbitrary dimension and spin, which do \emph{not} carry any energy-momentum. These primaries can either be thought of as living on the celestial sphere, or, on a spherical cross-section of the null-cone in momentum space. One may be tempted to call these primaries "soft operators" \cite{Strominger:2013lka, Strominger:2013jfa, He, Strominger:2014pwa, He:2014cra, Kapec:2015ena,Kapec:2014zla,Kapec:2015vwa,Pate:2017fgt, Campiglia:2015qka, Campiglia:2015kxa,Strominger:2017zoo,Kapec:2016jld, Kapec:2017gsg,Banerjee:2018fgd,Donnay:2018neh}, but let us emphasise that at this stage we \emph{cannot} identify them with \emph{the soft operators which appear in the soft-theorems}. In fact, this identification is one of  the problems that we try to solve in this paper. In any case, for simplicity of notation \emph{we will refer to these primaries as soft-operators but with the caveat, that they are not necessarily the same as the ones appearing in the soft theorems}. 





The next step is to write the Ward-identity in a useful form. In order to do this, we write the soft-charge $Q_S$, which appears in the Ward-identity, as a sum of conformal descendants of various primaries. A large portion of the paper is an attempt to show that this construction is \emph{useful}, i.e, the form of the "soft-charge", so constructed, is \emph{almost} uniquely determined by the structure of the Ward-identity and conformal invariance. Now to be completely honest, in this approach, one should also think of including the contribution of (hard) finite energy operators to the soft-charge and then show that \emph{such contributions are ruled out} by symmetry or by some more general principle. But we do not know how to systematically implement this.

So in this way we construct a theory in which the infinite dimensional global symmetries are built-in. But the main problem is that we do not yet know the dynamical principles which determine all the $S$--matrix elements in this theory, at least in principle. The only tool at our disposal is symmetry and it is clear that in general we will not be able to determine the $S$--matrix based on symmetry alone. In fact, in our case, the problem may be even more serious. It may happen that there exists no such theory in which the soft-charge $Q_S$ is \emph{actually} given by the expression we have. In other words, the theory, so constructed, is inconsistent or empty. But, somewhat surprisingly, \emph{conformal invariance} guarantees that this is not the case. What we find is that, with our choice of soft charge $Q_S$, conformal invariance alone allows us to (almost)\footnote{"Almost" because, at the end, we are not able to solve one technical problem related to the crossing symmetry of the $S$--matrix. We leave its solution to the experts.} \emph{solve the Ward-identity} and determine the $S$-matrix elements with the insertion of soft-operators. The solutions are given by soft-theorems. Although this is the final result, our primary motivation is to study the \emph{role of global symmetries} in this context. So let us now summarise the results :

We find that solvability of the Ward-identity depends crucially on \emph{the decoupling of primary descendants or null-states} of certain soft-operators. In \emph{even} space-time dimensions $D\ge 6$ the reason for the decoupling is (Lorentz) conformal invariance and the existence of the infinite dimensional global symmetry. In fact, the dimension and spin of the soft-operator and the null-state, which decouples, are \emph{completely determined by (Lorentz) conformal invariance and the type of infinite dimensional global symmetry}. Decoupling forces the corresponding soft-operator to satisfy linear PDE on the celestial sphere. Curiously, these PDEs are equations of motion of Euclidean gauge theories on the celestial sphere, with \emph{scalar} gauge-invariance. Now, the PDE satisfied by the soft-operator can be converted into PDE for the $S$--matrix elements with the insertion of the soft-operator. These equations can then  be solved subject to appropriate boundary conditions on the celestial sphere. Happily, these boundary conditions are also provided by (Lorentz) conformal invariance. The solutions have an interesting "pure-gauge" form. The role of this "pure-gauge" solution is to effectively reduce the number of degrees of freedom of the soft-operator. For example, in $D=5+1$ a leading soft-photon operator is a $(\D=1,l=1)$ primary with four degrees of freedom or polarization states. The solution of the differential equation shows that the four degrees of freedom are effectively replaced by \emph{one} scalar degree of freedom. Reduction in the effective number of degrees of freedom makes the solution of the Ward-identity \emph{almost} \footnote{"Almost" because, at the end, we are not able to solve one technical problem related to the crossing symmetry of the $S$--matrix. We leave its solution to the experts.} possible.

In $D=3+1$ decoupling of null-states is not required by (Lorentz) conformal invariance. But solvability of the Ward-identity requires such decoupling. We discuss this in detail in the paper.


\section{Notation and conventions}
For the convenience of the reader we collect some useful formulas here. For details we refer to \cite{Kapec:2017gsg,Pasterski:2017kqt}. 

In this paper we work exclusively with massless particles and \emph{even} space-time dimensions. In $D$ dimensional Minkowski space-time we parametrize the null momentum $p^{\mu}(\omega, \vec x)$ of a massless particle as, 
\be
p^{\mu} = \omega (1 + \vec x^2 , 2 \vec x, 1-\vec x^2), \quad  \vec x\in R^{D-2} = R^n
\ee
where $\omega$ is a real number. The Lorentz group $SO(D-1,1)=SO(n+1,1)$ acts on $R^n$ as the group of conformal transformations. The corresponding transformation of $\omega$ is given by, 
\be
\omega' = \bigg| \frac{\pa \vec x'}{\pa \vec x} \bigg| ^{-\frac{1}{n}} \omega
\ee
In the rest of the paper we omit the vector sign on $\vec x$ and simply denote it by $x$. 

Also in this paper we mostly work in $D=3+1$ or $n=2$ and $D=5+1$ or $n=4$.

\section{Infinite dimensional global U(1) symmetry}
Let us consider a theory of \emph{free} massless charged scalar fields. We are taking scalar field for simplicity but the following discussion has a straightforward extension to massless spinning charged particles. Now this theory has a $U(1)$ global symmetry. The corresponding charge can be written as, 
\be
Q_{0} = e \int d\mu(\omega, x) ( a^{\dagger}(\omega,x) a(\omega,x) - b^{\dagger}(\omega,x) b(\omega,x)) 
\ee
where $d\mu(\omega,\vec x)$ is the Lorentz invariant measure for massless particles written in terms of $(\omega, \vec x)$ and $(-e)e$ is the charge of the (anti-)particle.  $Q_0$ is a conserved charge. We can generalize this by defining $Q_{0}(f)$ as, 
\be
Q_{0}(f) = e \int d\mu(\omega, x) f(x) (a^{\dagger}(\omega,\vec x) a(\omega,\vec x) - b^{\dagger}(\omega,\vec x) b(\omega,\vec x))
\ee
where $f(x)$ is an arbitrary function. In the free theory $Q_0(f)$ is also a conserved quantity because the number operators $a^{\dagger}a, b^{\dagger}b$ are themselves conserved. This gives an infinite number of conserved charges corresponding to each function $f(x)$ \cite{Banerjee:2018fgd}. It acts on the creation operators as, 
\be
\begin{gathered}
e^{iQ_{0}(f)} a^{\dagger}(\omega,x) e^{-iQ_{0}(f)} = e^{i e f(x)} a^{\dagger} (\omega,x), \quad  e^{iQ_0(f)} \ket{\omega,x,e} = e^{ie f(x)} \ket{\omega,x,e} \\
e^{iQ_{0}(f)} b^{\dagger}(\omega,x) e^{-iQ_{0}(f)} = e^{-ie f(x)} b^{\dagger} (\omega,x), \quad  e^{iQ_0(f)} \ket{\omega,x,-e} = e^{-ie f(x)} \ket{\omega,x,-e}
\end{gathered}
\ee
So we can see that $Q_{0}(f)$ generates a $U(1)$ rotation at every point $x$ and the angle of rotation is determined by the value of the function $f$ at that point. \emph{We will refer to this infinite dimensional global $U(1)$ symmetry simply as $U(1)$. But it should be kept in mind that this is different from the global $U(1)$ symmetry whose angle of rotation does not depend on $x$}. 

Now using the Lorentz transformation property of the creation-annihilation operator it is easy to check that, 
\be\label{lf}
\boxed{
U(\Lambda) Q_{0}(f) U(\Lambda)^{-1} = Q_{0}(f'),  \quad  f'(x) = f(\Lambda^{-1} x)}
\ee
This transformation property of the conserved charge plays a central role in this paper. 

Now suppose we let the particles interact. The interacting theory has a non-trivial $S$-matrix and it is easy to see that the $U(1)$ cannot be a symmetry of the $S$-matrix. 

To see this, consider the (in) out states $\ket{\alpha, (in) out}$. Here $\alpha$ is a collective index for the momenta and charges of the initial or final state particles, i.e, $\alpha= \{\omega_i,x_i, q_i \}$. The $S$-matrix element for the reaction $\beta \rightarrow \alpha$ is given by the scalar product $S_{\alpha\beta} = \bra{\alpha,  out}\ket{\beta,  in}$. Now let $\tilde Q_0(f)$ be conserved charge which acts on \textit{both} the in and out states as,
\be\label{inout}
\begin{gathered}
\tilde Q_0(f) \ket{\alpha, in} = \bigg( \sum_{i\in\alpha} q_i f(x_i) \bigg) \ket{\alpha, in} \\
\bra{\alpha, out} \tilde Q_0(f) = \bra{\alpha, out} \bigg( \sum_{i\in\alpha} q_i f(x_i) \bigg) 
\end{gathered}
\ee
Now since $\tilde Q_0(f)$ is Hermitian, we get

\be\label{wi}
\bigg(\sum_{i\in\alpha} q_i f(x_i) - \sum_{i\in\beta} q_i f(x_i) \bigg) \bra{\alpha, out}\ket{\beta, in} = \bigg(\sum_{i\in\alpha} q_i f(x_i) - \sum_{i\in\beta} q_i f(x_i) \bigg) S_{\alpha\beta} = 0
\ee

Now it is clear that this identity cannot hold if we take $f(x)$ to be arbitrary and $S_{\alpha\beta} \ne 0$ for generic values of incoming and outgoing momenta. For example, we can take $f$ to be non-vanishing only at the location of, say, the $l$-th charged particle and then \eqref{wi} implies that $S_{\alpha\beta}=0$ because $q_l\ne0$. Therefore $\tilde Q_0(f)$ cannot be a conserved charge in the interacting scalar theory.

Now it is a remarkable fact \cite{Strominger:2013lka, He:2014cra, Kapec:2015ena,Kapec:2014zla,Campiglia:2015qka, Strominger:2017zoo} that if we add photons to the interacting theory then one can add correction terms to $\tilde Q_0(f)$ and the corrected charge $Q(f)$ commutes with the $S$-matrix. In this case the total charge can be written as, $Q(f) = Q_H(f) + Q_S(f)$. $Q_H(f)$ is called the hard-charge and is required to generate the $U(1)$ transformation on the charged particles, i.e,
\be\label{hc}
\begin{gathered}
Q^{in}_{H}(f) \ket{\alpha, in} = \bigg( \sum_{k\in \alpha} q_k f(x_k)\bigg) \ket{\alpha, in} \\
\bra{\beta, out} Q^{out}_{H}(f) = \bra{\beta, out} \bigg( \sum_{k\in \beta} q_k f(x_k)\bigg)
\end{gathered}
\ee
We have written $Q^{in/out}_H(f)$ because $Q_H(f)$ by itself is not conserved and so $Q^{in}_H(f)\ne Q^{out}_H(f)$. The second part $Q_S(f)$ is called the soft-charge and is constructed out of soft (zero energy) photons. 
 
The statement of conservation is,
 \be
 Q(f) = Q^{in}_H(f) + Q^{in}_S(f) = Q^{out}_H(f) + Q^{out}_S(f)
  \ee
 This can be written in the form of a Ward-identity \cite{Strominger:2013lka, He:2014cra, Kapec:2015ena,Kapec:2014zla,Campiglia:2015qka, Strominger:2017zoo} as, 
 \be\label{u1}
 \boxed{
 \bra{\beta, out} Q^{out}_S(f)\ket{\alpha, in}  - \bra{\beta, out} Q^{in}_S(f)\ket{\alpha, in} = \bigg( \sum_{i\in\alpha} q_i f(x_i) - \sum_{i\in\beta} q_i f(x_i) \bigg) \bra{\beta, out}\ket{\alpha, in}}
  \ee
This Ward-identity \eqref{u1} (and a similar one for the supertranslation) is the starting point of our paper. In other words, instead of deriving it in a specific theory we \emph{assume} that this holds in our theory.  If the symmetries are powerful enough they will determine the theory to a large extent.  

\section{Lorentz transformation of the soft-charge : A necessary condition} 
We want to find out the Lorentz transformation property of the soft-charge $Q_S(f)$. For notational simplicity let us define, 
\be
\Lambda\alpha := \{ \Lambda \omega_i,\Lambda x_i, q_i\}
\ee
where $\Lambda$ is a Lorentz transformation. 

Let us now apply Lorentz transformation $U(\Lambda)$ to both sides of the Ward-identity \eqref{u1}. This gives us,
\be
\begin{gathered}
 \bra{\Lambda\beta, out} U(\Lambda)Q^{out}_S(f)U(\Lambda)^{\dagger}\ket{\Lambda\alpha, in}  - \bra{\Lambda\beta, out} U(\Lambda) Q^{in}_S(f) U(\Lambda)^{\dagger}\ket{\Lambda\alpha, in} \\
= \bigg( \sum_{i\in\alpha} q_i f(x_i) - \sum_{i\in\beta} q_i f(x_i) \bigg) \bra{\Lambda\beta, out}\ket{\Lambda\alpha, in}
\end{gathered}
\ee
Now we make a change of variable $(\omega,x)\rightarrow(\Lambda^{-1}\omega,\Lambda^{-1}x)$ and this gives us,
\be\label{lt}
\begin{gathered}
 \bra{\beta, out} U(\Lambda)Q^{out}_S(f)U(\Lambda)^{\dagger}\ket{\alpha, in}  - \bra{\beta, out} U(\Lambda) Q^{in}_S(f) U(\Lambda)^{\dagger}\ket{\alpha, in} \\
= \bigg( \sum_{i\in\alpha} q_i f(\Lambda^{-1} x_i) - \sum_{i\in\beta} q_i f(\Lambda^{-1} x_i) \bigg) \bra{\beta, out}\ket{\alpha, in} \\
= \bra{\beta, out} Q^{out}_S(f')\ket{\alpha, in}  - \bra{\beta, out} Q^{in}_S(f') \ket{\alpha, in}
\end{gathered}
\ee
where we have defined,
\be
f'(x) = f(\Lambda^{-1}x)
\ee
We can write \eqref{lt} as, 
\be
\begin{gathered}
\bra{\beta, out} U(\Lambda)Q^{out}_S(f)U(\Lambda)^{\dagger} -Q^{out}_S(f') \ket{\alpha, in}  \\
= \bra{\beta, out} U(\Lambda) Q^{in}_S(f) U(\Lambda)^{\dagger} - Q^{in}_S(f') \ket{\alpha, in}
\end{gathered}
\ee
This tells that there are infinite number of conserved charges (Ward-identities) depending on an arbitrary function $f(x)$ and an arbitrary Lorentz transformation $\Lambda$. \emph{We simply assume that no such charge or Ward-identity exists}, i.e, 
\be
\boxed{
\begin{gathered}
U(\Lambda)Q^{out}_S(f)U(\Lambda)^{\dagger}  = Q^{out}_S(f') \\
U(\Lambda) Q^{in}_S(f) U(\Lambda)^{\dagger} = Q^{in}_S(f') 
\end{gathered} }
\ee
This gives us the Lorentz transformation property of the soft-charge.

There is another way to see this if we make the following assumption : 

\emph{The Lorentz transformation property of the \underline{conserved charge} $Q(f)$ in the interacting theory is the same as the Lorentz transformation property of the \underline{conserved charge} $Q_0(f)$ in the free theory}. This seems to be a reasonable assumption. 

Using this assumption and the Lorentz transformation property of the free charge $Q_0(f)$ given by \eqref{lf}, we get, 
\be\label{ct}
U(\Lambda) Q(f) U(\Lambda)^{-1} = Q(f') , \quad  f'(x) = f(\Lambda^{-1}x)
\ee
Let us now find out the Lorentz transformation property of the hard-charge $Q_H(f)$. $Q_H(f)$ generates $U(1)$ transformation on charged particles as given by \eqref{hc}. Lorentz transforming both sides of this equation we get, 
\be
U(\Lambda)Q_{H}(f) U(\Lambda)^{-1} U(\Lambda) \ket{\alpha} = \big( \sum_k q_k f(x_k)\big) U(\Lambda)\ket{\alpha}
\ee
$\Rightarrow$
\be
U(\Lambda)Q_{H}(f) U(\Lambda)^{-1}\ket{\Lambda\alpha} = \big( \sum_k q_k f(x_k)\big) \ket{\Lambda\alpha}
\ee
Now replacing $(\omega,x)$ with $(\Lambda^{-1}\omega,\Lambda^{-1}x)$ we get, 
\be
U(\Lambda)Q_{H}(f) U(\Lambda)^{-1}\ket{\alpha} = \big( \sum_k q_k f(\Lambda^{-1}x_k)\big) \ket{\alpha} = Q_H(f')\ket{\alpha} , \quad  f'(x) = f(\Lambda^{-1}x)
\ee
So we can write,
\be\label{ht}
U(\Lambda) Q_H(f) U(\Lambda)^{-1} = Q_H(f') , \quad  f'(x) = f(\Lambda^{-1}x)
\ee
Now since $Q(f)= Q_H(f) + Q_S(f)$, using \eqref{ct} and \eqref{ht} we get, 
\be\label{sl}
\boxed{
U(\Lambda) Q_S(f) U(\Lambda)^{-1} = Q_S(f'),  \quad  f'(x) = f(\Lambda^{-1}x)}
\ee
This is the Lorentz transformation property of the soft-charge $Q_S(f)$ which we also obtained from the Ward-identity.

We now write $Q_S(f)$ as the integral of a local operator $O(x)$ weighted by $f(x)$ \footnote{We can write this because it follows from the Ward-identity \eqref{u1} that $Q_S(f)$ is a linear (operator-valued) functional of $f$, i.e, $Q_S(\alpha f + g) = \alpha Q_S(f) + Q_S(g)$ for any $\alpha \in \mathbb{C}$. Now if we think of $f$ as a vector $\ket{f}$ in an infinite dimensional Hilbert space then we can write, $Q_S(\ket{f}) = Q_S (\int d^nx f(x) \ket{x}) = \int d^nx f(x) Q_S(\ket{x})= \int d^nx f(x) O(x)$. Here we have defined $O(x_0) = Q_S(\ket{x_0}) \equiv Q_S(f(x)=\delta^n(x-x_0))$.}, i.e, 
\be
Q_S(f) = \int d^nx f(x) O(x)
\ee

Let us now evaluate $Q_S(f')$ where $f'(x)=f(\Lambda^{-1}x)$. It is given by,
\be\label{e2}
Q_S(f')=\int d^n x f'(x) O(x) = \int d^n x f(\Lambda^{-1}x) O(x) = \int d^n (\Lambda x) f(x) O(\Lambda x) = \int d^n x f(x) \bigg|\frac{\partial x'}{\partial x}\bigg |O(x')
\ee
where $x'=\Lambda x$.  


From \eqref{sl} we also have, 
\be\label{e1}
Q_S(f') = U(\Lambda) Q_S(f) U(\Lambda)^{-1} = \int d^n x f(x) U(\Lambda) O(x) U(\Lambda)^{-1}
\ee
So comparing \eqref{e2} and \eqref{e1} we get, 
\be\label{ii}
 \int d^n x f(x) U(\Lambda) O(x) U(\Lambda)^{-1} = \int d^n x f(x) \bigg|\frac{\partial x'}{\partial x}\bigg |O(x')
\ee
Since this is true for \textit{any} function $f(x)$ we get the \textit{local} transformation law,
\be\label{li}
\boxed{U(\Lambda) O(x) U(\Lambda)^{-1} = O'(x) = \bigg|\frac{\partial x'}{\partial x}\bigg |O(x') = \bigg|\frac{\partial x'}{\partial x}\bigg |^{\frac{\Delta}{n}} O(x')}  \quad  x' = \Lambda x, \quad  \Delta = n
\ee
Now as Lorentz transformation acts on the $x\in R^n$ coordinates as conformal transformations, this shows that the operator \emph{$O(x)$ is a scalar conformal primary of weight $\Delta =n$}.


\eqref{li} is a \textit{necessary} condition which needs to be satisfied by \textit{any} potential candidate for the operator $O(x)$. We now have to find some way of constructing candidates for $O(x)$.  
 
\subsection{Operator content}
In order to construct possible candidates for $O(x)$ we need to know the operator content of the theory. A subset of these operators can be taken to be the momentum-space creation-annihilation operators which create asymptotic physical states with finite energy. Now, an important lesson following from our current understanding of soft-theorem -- Ward identity correspondence is that we should also add to this list the \textit{soft operators} $S_{\alpha}(x)$ \cite{Strominger:2013lka, Strominger:2013jfa, He, Strominger:2014pwa, He:2014cra, Kapec:2015ena,Kapec:2014zla,Kapec:2015vwa,Pate:2017fgt, Campiglia:2015qka, Campiglia:2015kxa,Strominger:2017zoo,Kapec:2016jld, Kapec:2017gsg,Banerjee:2018fgd,Donnay:2018neh}. They are usually defined as, 
\be
S_{\alpha}(x) = \displaystyle\lim_{\omega \to 0} D(\omega) A(\omega,x,\alpha)
\ee
where $A(\omega,x,\alpha)$ is the creation/annihilation operator of a massless particle of helicity $\alpha$ and $D(\omega)$ is a differential operator built out of energy $\omega$ and $\pa_{\omega}$. The helicity index $\alpha$ belongs to an irreducible representation of the little group $SO(n)$. For a massless particle of integer spin $l$ the relevant representation is the rank-$l$ symmetric traceless tensor of $SO(n)$. For our purpose, the most important point is that : \emph{there exist non-trivial operators $S_{\alpha}(x)$ which under (Lorentz) conformal transformation transform as \underline{primary} operators}. We can summarise this observation in the form of an assumption :

\underline{Assumption}: \emph{There are an infinite number of operators denoted by $S^{\Delta}_{a_1a_2...a_l}(x)$, not all of which are trivial and which transform under (Lorentz) conformal transformation as a primary operator of weight $\Delta$ and spin $l$. We also add to this list all the conformal descendants of all the primaries. So each $S^{\Delta}_{a_1a_2...a_l}(x)$ together with its descendants form a complete representation of the conformal group $SO(n+1,1)$. We further assume that the primary operators $S^{\Delta}_{a_1a_2...a_l}(x)$ and their descendants carry zero energy-momentum.} 

Here $a$ is the vector index of $SO(n)$. 


Let us now make few comments on this assumption. First of all, as we have already mentioned in the introduction, the primaries $S^{\Delta}_{a_1a_2...a_l}(x)$ are \emph{not necessarily the same as the soft operators which appear in the soft-theorems}. For our purpose, the assumption stated above completely characterises the operators $S^{\Delta}_{a_1a_2...a_l}(x)$. Now there is some flexibility in the choice of the operator basis. For example, if we have both $S^1_a(x)$ and $S^0(x)$ then we can construct another $(\D=1,l=1)$ primary given by, $\tilde S^1_a(x) = S^1_a(x) + \alpha \pa_a S^0(x)$, where $\alpha$ is an arbitrary real numbers. This is the simplest example but there are many more. This \emph{freedom of basis redefinition} plays an interesting role in the later part of the paper. 

The second part of the assumption which says that the primary operators and their descendants do not carry energy-momentum has the following consequence. Suppose $\bra{\{\omega_i,x_i,out\}} S^1_{a, in/out}(x)\ket{\{\omega_j,x_j,in\}}$ is a $S$--matrix element with an insertion of the primary $S^1_a(x)$. Then this assumption tells us that the location of the primary, given by $x$, can be varied \emph{independently} of the remaining $\{\omega_k,x_k\}$. This is a very useful fact which we use throughout the paper. 


The second assumption is that : \emph{The operator $O(x)$ is either a primary by itself or a \underline{primary descendent} of another primary $S^{\D}_{a_1a_2\cdots a_l}$ or a sum of \underline{primary descendants} of more than one $S^{\D}_{a_1a_2\cdots a_l}$.} As we have already discussed in the introduction, this is a useful assumption. We will show that the operator $O(x)$, so constructed, is almost uniquely determined by conformal invariance. 



Let us now discuss that why we expect (conformal) Lorentz invariance to be a non--trivial constraint in this setting. 

 In hindsight we know that $O(x)= (\partial^2)^{\frac{n-2}{2}} \partial_a S^1_a(x)$ \cite{Kapec:2014zla} \footnote{There may be a proportionality constant between $O(x)$ and $(\partial^2)^{\frac{n-2}{2}} \partial_a S_a(x)$. We neglect this for simplicity because we are concerned only with the Lorentz transformation property of the operator $(\partial^2)^{\frac{n-2}{2}} \partial_a S_a(x)$.} where $S^1_a(x)$ is the leading soft-photon operator which is a conformal primary with $(\Delta=1, l=1)$. Now we also know from \eqref{li} that $O(x)$ must be a primary with $(\Delta=n, l=0)$. This tells us that \textit{$S^1_a(x)$ must have a scalar descendent at level $(n-1)$ \textit{which is also a primary}}. This is a non-trivial constraint because not every $(\Delta=1,l=1)$ primary has a scalar primary descendent at level $(n-1)$. We show that \textit{for $n=4$ this is possible only if the primary descendent $\pa_a(\pa_aS^1_b-\pa_b S^1_a)$ decouples from the $S$-matrix}. This leads to a differential equation for the $S$-matrix element with an insertion of $S^1_a(x)$. The same phenomenon happens in the case of soft-graviton. Now we generalise this in the following way. We replace $S^1_a(x)$ with an arbitrary primary operator $S^{\Delta}_{a_1...a_l}$ with dimension $\Delta$ and integer spin $l$ which has a scalar descendent $O(x)$ at level $(n-\Delta)$. This requires $\Delta\le n$. Now there are infinite number of such operators. Consider, for example, the $(\Delta=1,l=3)$ primary operator $S^1_{abc}(x)$. This can be a leading soft spin-$3$ operator. This has a scalar descendent $O(x) = (\pa^2)^{\frac{n-4}{2}}\pa_a\pa_b\pa_c S^1_{abc}(x)$ of dimension $\Delta=n$. We show that at least for $n=4$, $O(x)=\pa_a\pa_b\pa_c S^1_{abc}(x)$ \textit{cannot be a primary operator}. So this is ruled out, i.e, \textit{ for $n=4$ or space-time dimension $D=6$, $S^1_{abc}(x)$ cannot contribute to the soft-charge of the $U(1)$ symmetry}. 
 
 \section{Potential candidates for $O(x)$ in space-time dimension $D=6$ or $n= D-2 =4$}
 
As we have described we want to consider different possibilities for $O(x)$. For the time being let us further impose the constraint $\Delta \ge 0$. We will come back to operators with $\Delta < 0$ later.  

\textit{A potential candidate for $O(x)$ is a ($\Delta=n=4,l=0$) \underline{primary} constructed from} $S^{\Delta}_{a_1a_2....a_l}$. 
Let us start with a ($ \D = 1,l=1 $) primary operator $S^1_a(x)$. Since the dimension of the operator is $ 1 $, we need at least three derivatives to make it dimension $ 4 $. The resulting descendent should also be a scalar. For $ l = 1 $ the only possibility is $ \pa^2 \pa_{a} S^{1}_{a}(x) $. For $(\D=1,l=2)$ one can check that there is no way to construct a ($ \D = 4, l = 0 $) descendent without considering fractional Laplacian. 

Similarly one can consider other possibilities. Below we have listed the ($\Delta\ge 0, l$) primary operators which have ($ \D = 4, l = 0 $) descendent :
\bea\label{poss_op_u1}
\no \( \D=0 \to \D = 4, \, l=0 \) &:& \hspace{10mm} O(x) = (\pa^2)^2 S^{0}(x), \hspace{3mm} \pa^2 \pa_a \pa_b S^{0}_{ab}(x), \hspace{3mm} \pa_a \pa_b \pa_c \pa_d S^{0}_{abcd}(x)\\
\no \( \D = 1 \to \D = 4, \, l=0 \) &:& \hspace{10mm} O(x) = \pa^2 \pa_a S^{1}_a(x) , \hspace{4mm} \pa_a \pa_b \pa_c S^{1}_{abc}(x) \\
\no \( \D = 2 \to \D = 4, \, l=0 \) &:& \hspace{10mm} O(x) = \pa^2 S^{2}(x), \hspace{8mm} \partial_a \partial_b S^{2}_{ab}(x) \\
\no \( \D = 3 \to \D = 4, \, l=0 \) &:& \hspace{10mm} O(x) = \pa_a S^{3}_{a}(x)
\eea

Our job is to check which of these operators can be primary. Let us start with the operator $O(x)=\pa^2 \pa_{a} S^{1}_{a}(x)$. 

\subsection{Operator Decoupling}
Let us denote the primary operator $S^{1}_a(x)$ simply by $S_a(x)$. 

Under (Lorentz) conformal transformation $\Lambda$, $S_a(x)$ transforms as, 
\be
S'_a(x') = \bigg|\frac{\partial x'}{\partial x}\bigg| ^{-\frac{1}{4}} R_{ab}(x) S_b(x) , \qquad  x' = \Lambda x
\ee
where $R_{ab}(x)$ is the local rotation matrix associated with the conformal transformation. 

Now consider an infinitesimal special conformal transformation (SCT) :
\be
\label{inf_sct}
x^a \to {x'}^a = x^a  - 2 \, (\e \cdot x) x^a + \e^a x^2
\ee
where $ \e^a $ are the infinitesimal parameters. Under this transformation we have
\be
\left| \f{\pa x'}{\pa x} \right| = 1 - 8 \, \e \cdot x 
\ee
and
\be
R_{ab}(x) = \delta_{ab} + 2 (\e_a x_b - \e_b x_a) = \d_{ab} + 2 \, \Om_{ab}(x) , \qquad  \Omega_{ab} = \e_a x_b - \e_b x_a
\ee
So the infinitesimal transformation of $S_a(x)$ can be written as,
\bea\label{spt}
\no S'_a(x') &=& \left| \f{\pa x'}{\pa x} \right|^{-\f{1}{4}} R_{ab}(x) S_b(x) \\
&=& (1 + 2\e \cdot x) S_a(x) + 2 \, \Om_{ab}(x)S_b(x) 
\eea

Let us now consider the operator $O(x)=\partial^2\partial_a S_a(x)$. It follows from \eqref{spt} that under infinitesimal SCT the operator $O(x)$ transforms as :
\be 
\label{o1}
O'(x') = \left( 1 + 8 \, \e \cdot x \right) O(x) + \boxed{4 \, \e_a \, \partial_b F_{ab}(x)}
\ee
where  we have defined,
\be 
F_{ab}(x) = \pa_a S_b(x) - \pa_b S_a(x)
\ee
Now we impose the constraint that the operator $O(x)$ must transform like a $(\Delta=4,l=0)$ primary. The first term in \eqref{o1} gives the standard transformation of a $(\D=4,l=0)$ primary. So if we want $O(x)=\partial^2\partial_a S_a(x)$ to be primary then we have to set the additional piece $\partial_a F_{ab}(x)$ to zero. Now \textit{the equation $\pa_a F_{ab}(x)=0$ is consistent or conformally invariant only if $\partial_a F_{ab}(x)$ itself is a primary operator.} One can easily check that this is indeed the case, i.e, $\partial_a F_{ab}(x)$ is a $(\D=3,l=1)$ primary. So we can set, 
\be\label{smc11}
\boxed{
\pa_a F_{ab} = \pa_a (\pa_a S_b - \pa_b S_a) =0 }
\ee 
Therefore we can see that the operator $O(x)=\partial^2\partial_a S_a(x)$ is a $(\Delta=4, l=0)$ primary only if $S_a(x)$ satisfies the constraint \eqref{smc11}.

\subsection{Differential Equation For S-matrix element}

Consider the S-matrix element with an insertion of $S_a(x)$ in the (incoming) outgoing channel, i.e, $\bra{\{\omega_i,x_i,q_i,out\}} S^{in/out}_a(x)\ket{\{\omega_j,x_j,q_j,in\}}$. Since the following calculation does not depend on the channel, for notational simplicity, we omit the (in) out superscript from $S_a(x)$. So the $S$--matrix element is now simply denoted by, $\bra{\{\omega_i,x_i,q_i,out\}} S_a(x)\ket{\{\omega_j,x_j,q_j,in\}}$. We also take all the incoming and outgoing particles to be scalars. 

Let us now define 
\be\label{vp}
A_a(x,\{\omega_\alpha, x_\alpha,q_\alpha\}) = \bra{\{\omega_i,x_i,q_i,out\}} S_a(x)\ket{\{\omega_j,x_j,q_j,in\}}
\ee
where the index $\alpha$ runs over both the incoming and outgoing particles. Now using the constraint equation \eqref{smc11}  we get, 
\be\boxed{
\bra{\{\omega_i,x_i,q_i,out\}} \pa_a (\pa_a S_b - \pa_b S_a)(x) \ket{\{\omega_j,x_j,q_j,in\}}} = 0
\ee
Since there is no ordering between the $(x, \{x_i,out\},\{x_j,in\})$ coordinates we can pull the derivates outside the $S$-matrix without producing contact-terms. This point probably requires further justification but for now we assume this to be true. The main justification for this is that the result obtained by assuming this is consistent with Weinberg's soft-photon theorem. Now using the definition \eqref{vp} we can write, 
\be\label{EM}
\pa_a(\pa_a A_b - \pa_b A_a ) = \pa_a F_{ab} =0 
\ee
where we have defined $F_{ab} = \pa_a A_b - \pa_b A_b$. Equation \eqref{EM} together with the Bianchi identity ($dF=0$) are the Maxwell equations on $R^4$. Now to solve equation \eqref{EM} we need boundary condition. The boundary condition can be obtained in the following way.


From the Lorentz invariance of the $S$ matrix we can write, 
\be\label{LT}
\bra{\{\omega'_i,x'_i,q_i,out\}} S'_a(x)\ket{\{\omega'_j,x'_j,q_j,in\}} = \bra{\{\omega_i,x_i,q_i,out\}} S_a(x)\ket{\{\omega_j,x_j,q_j,in\}}\ee

where 
\be
\label{fields}
S'_a(x) = \bigg|\frac{\partial x'}{\partial x}\bigg| ^{\frac{1}{4}} R^{-1}_{ab}(x) S_b(x') = \bigg|\frac{\partial x'}{\partial x}\bigg| ^{\frac{1}{4}} R_{ba}(x) S_b(x')
\ee
Here we have used the fact that $R$ is an orthogonal matrix. Now consider an inversion about the origin given by, 
\be
X_a \rightarrow X'_a = \frac{X_a}{X^2}
\ee
In this case, 
\be
\bigg|\frac{\partial x'}{\partial x} \bigg| = \frac{1}{(x^2)^4} 
\ee
and
\be
R_{ab}(x) = R_{ba}(x) = \delta_{ab} - 2 \frac{x_a x_b}{x^2} = I_{ab}(x)
\ee
Therefore under inversion, 
\be
S'_a(x) = \frac{1}{x^2} \bigg(\delta_{ab} - 2 \frac{x_ax_b}{x^2}\bigg) S_b(x') = \frac{1}{x^2} I_{ab}(x) S_b(x') , \quad x'_a = \frac{x_a}{x^2}
\ee
Now using \eqref{LT} we get, 
\be
\bra{\{\omega_i,x_i,q_i,out\}} S_a(x)\ket{\{\omega_j,x_j,q_j,in\}} = \frac{1}{x^2} I_{ab}(x) \bra{\{\omega'_i,x'_i,q_i,out\}} S_a(x')\ket{\{\omega'_j,x'_j,q_j,in\}}
\ee

Let us assume that none of the $\{x_i,out\}$ or $\{x_j,in\}$ is at $\infty$. Now we take $x\rightarrow \infty$ keeping all the incoming and outgoing momenta fixed. In this limit $x'_a = x_a / x^2 \rightarrow 0$ and so the leading term in a $1/x$ expansion of the $S$-matrix element is given by, 
\be\label{moff}
\boxed{
\bra{\{\omega_i,x_i,q_i,out\}} S_a(x)\ket{\{\omega_j,x_j,q_j,in\}} \xrightarrow{\text{$x\rightarrow\infty$}}  \frac{1}{x^2} I_{ab}(x) M_b(\{\omega_\alpha,x_\alpha,q_\alpha\}) + O(\frac{1}{x^3})}
\ee
where
\be 
M_a(\{\omega_\alpha, x_\alpha,q_\alpha\}) = \bra{\{\omega'_i,x'_i,q_i,out\}} S_a(0)\ket{\{\omega'_j,x'_j,q_j,in\}}
\ee
This is expected to be finite because according to our assumption, none of the $x_\alpha$ is at $\infty$ and so the image $x'_\alpha$ under inversion $\ne 0$. The important point is that $M_a(\{\omega_\alpha, x_\alpha,q_\alpha\})$ does not depend on $x$. Also note that $I_{ab}(x)$ is $O(1)$ as $x\rightarrow\infty$.  

In terms of $A_a$ the fall-off condition \eqref{moff} can be rewritten as, 
\be\label{vpf}
\boxed{
A_a(x,\{\omega_\alpha, x_\alpha ,q_\alpha\}) \xrightarrow{\text{$x\rightarrow\infty$}}  \frac{1}{x^2} I_{ab}(x) M_b(\{\omega_\alpha,x_\alpha,q_\alpha\}) + O(\frac{1}{x^3})}
\ee
The reader may be skeptical about the use of inversion. So as a consistency check, we show in section \eqref{inv} that the boundary condition \eqref{vpf}, together with Weinberg's soft-photon theorem, is equivalent to conservation of charge.  

Let us now solve the Maxwell's equation \eqref{EM}. Since we are in Euclidean space, instead of the wave equation, the Electric fields $E_i ( = F_{i4})$, and the magnetic fields $B_i ( = \frac{1}{2} \epsilon_{ijk} F_{jk} )$ now satisfy the \textit{four dimensional Laplace's} equation, 
\be
\pa_a \pa_a E_i = \pa_a \pa_a B_i = 0 
\ee

This, together with the falloff condition $F_{ab}\sim O(\frac{1}{x^3})$ as $x\rightarrow\infty$ -- derived from the fall-off condition \eqref{vpf} of $A_a$ -- implies that $E_i = B_i = 0$. Here we have used the fact that a function which is harmonic everywhere and vanishes at infinity is identically zero.

 Therefore $F_{ab}=0$ and we can write, 
\be\label{scu}
\boxed{
\pa_a A_b - \pa_b A_a =0 \Longleftrightarrow A_a(x,\{\omega_\alpha, x_\alpha,q_\alpha\}) = \pa_a \Lambda(x, \{\omega_\alpha,x_\alpha,q_\alpha\})} 
\ee
Therefore \textit{the four soft S-matrices $A_a$ corresponding to four helicity states of the photon are determined in terms of a single scalar function $\Lambda(x,\{\omega_\alpha,x_\alpha,q_\alpha\})$}. To be more precise, we can write,
\be
\bra{\{\omega_i,x_i,q_i,out\}} S^{out}_a(x)\ket{\{\omega_j,x_j,q_j,in\}} = \pa_a \Lambda_{out}(x,\{\omega_\alpha,x_\alpha,q_\alpha\})
\ee
and
\be
\bra{\{\omega_i,x_i,q_i,out\}} S^{in}_a(x)\ket{\{\omega_j,x_j,q_j,in\}} = \pa_a \Lambda_{in}(x,\{\omega_\alpha,x_\alpha,q_\alpha\})
\ee

This result is actually consistent with Weinberg's soft-photon theorem \cite{Weinberg:1965nx}. To see this let us first write the soft-photon theorem, say for an outgoing soft photon, as \cite{Kapec:2017gsg,Kapec:2014zla}, 
\be
\begin{gathered}
\bra{\{\omega_i,x_i,q_i,out\}} S^{out}_a(x)\ket{\{\omega_j,x_j,q_j,in\}}  \\
= \gamma \bigg( \sum_{i\in out} q_i \frac{2 (x-x_i)_a}{(x-x_i)^2} - \sum_{j\in in} q_j \frac{2 (x-x_j)_a}{(x-x_j)^2} \bigg)  \bra{\{\omega_i,x_i,q_i,out\}}\ket{\{\omega_j,x_j,q_j,in\}}
\end{gathered}
\ee
where $\gamma$ is a numerical constant whose precise value is not important for us. Now it is easy to see that, $\Lambda_{out}$ is given by,
\be
\Lambda_{out}(x, \{\omega_\alpha,x_\alpha,q_\alpha\}) = \gamma \bigg(\sum_{i\in out} q_i \ln (x-x_i)^2 - \sum_{j\in in} q_j \ln (x-x_j)^2 \bigg) \bra{\{\omega_i,x_i,q_i,out\}}\ket{\{\omega_j,x_j,q_j,in\}}
\ee


\subsection{Other operators in the list}
Let us now consider the other operators in the list of potential candidates for $O(x)$ subject to the constraint $\D\ge 0$.
\bea
\no \( \D=0 \to \D = 4, \, l=0 \) &:& \hspace{10mm} O(x) = (\pa^2)^2 S^{0}(x), \hspace{3mm} \pa^2 \pa_a \pa_b S^{0}_{ab}(x), \hspace{3mm} \pa_a \pa_b \pa_c \pa_d S^{0}_{abcd}(x)\\
\no \( \D = 1 \to \D = 4, \, l=0 \) &:& \hspace{10mm} O(x) = \pa^2 \pa_a S^{1}_a(x) , \hspace{4mm} \pa_a \pa_b \pa_c S^{1}_{abc}(x) \\
\no \( \D = 2 \to \D = 4, \, l=0 \) &:& \hspace{10mm} O(x) = \pa^2 S^{2}(x), \hspace{8mm} \partial_a \partial_b S^{2}_{ab}(x) \\
\no \( \D = 3 \to \D = 4, \, l=0 \) &:& \hspace{10mm} O(x) = \pa_a S^{3}_{a}(x)
\eea

Although the above operators have the required scaling dimension and spin, most of them are not primary or cannot be made primary by imposing additional constraints on $S^{\Delta}_{a_1a_2...a_l}$. For example, let us consider the operator $ O(x)= \pa_a \pa_b \pa_c S^{1}_{abc}(x) $. Since $S^1_{abc}(x)$ is a $(\Delta=1,l=1)$ primary it transforms under infinitesimal SCT \eqref{inf_sct} as,
\be
\no S'^{1}_{abc}(x') = \( 1 + 2 \, \e \cdot x \) S^{1}_{abc}(x) + 2 \, \Om_{ad} S^{1}_{dbc}(x) + 2 \, \Om_{bd} S^{1}_{adc}(x) + 2 \, \Om_{cd} S^{1}_{abd}(x)
\ee
Therefore,
\be
\begin{gathered}
\no O'(x') =  \partial'_a \partial'_b \pa'_c S'^{1}_{abc}(x') = \( 1 + 8 \, \e \cdot x \)\partial_a \partial_b \pa_c S^{1}_{abc}(x) - 12 \,\e_c \pa_a \, \pa_b \, S^{1}_{abc}(x) \\
 = ( 1 + 8 \, \e \cdot x \ ) O(x) - \boxed{12 \,\e_c \pa_a \, \pa_b \, S^{1}_{abc}(x)}
\end{gathered}
\ee
Now if we want $O(x)$ to be $(\Delta=4,l=0)$ primary then we should set the boxed operator $\pa_a\pa_b S^1_{abc}$ to zero. This is consistent if $\pa_a\pa_b S^1_{abc}$ is a primary operator. But this is not a primary operator. In fact,
\be
 \no \partial'_a \pa'_b S'^{1}_{abc}(x') = \( 1 + 6 \, \e \cdot x \)\partial_a \partial_b  S^{1}_{abc}(x) + 2 \, \Om_{cd} \partial_a \partial_b S^{1}_{abd}(x) - \boxed{12 \,\e_a  \, \pa_b \, S^{1}_{abc}(x)}
\ee
and
\be
\no \partial'_a S'^{1}_{abc}(x') = \( 1 + 4 \, \e \cdot x \)\partial_a S^{1}_{abc}(x) + 2 \, \Om_{bd} \pa_a S^{1}_{adc}(x) + 2 \, \Om_{cd} \pa_a S^{1}_{abd}(x) -  \boxed{8 \,\e_a  \, S^{1}_{abc}(x)}
\label{p2a2}
\ee
Thus to make $O(x)=\pa_a \pa_b \pa_c S^{1}_{abc}(x)$ a $(\D=4,l=0)$ primary we have to set $ S^{1}_{abc}(x) = 0 $ which is of course the trivial solution. 

Similarly one can consider the other operators in the above list and check that only the operators $ (\pa^2)^2 S^{0}(x)$ and $\pa_a S^{3}_{a}(x) $ are primary. The transformation properties of the operators $ (\pa^2)^2 S^{0}(x)$ and $\pa_a S^{3}_{a}(x)$ under infinitesimal SCT \eqref{inf_sct} are,
\be
\label{firstp}
(\pa'^2)^2 S'^{0}(x') = (1 + 8 \, \e \cdot x) \, (\pa^2)^2 \, S^{0}(x)
\ee
and
\be
\label{lastp}
\pa'_a S'^{3}_{a}(x') = (1 + 8 \, \e \cdot x) \, \pa_a \, S^{3}_{a}(x)
\ee

Therefore these two operators are $ (\D=4, \, l=0)$ primaries without any constraint. 


Therefore from our analysis we can see that if we restrict our attention to primary operators $S^{\Delta}_{a_1a_2...a_l}$ with $\Delta\ge 0$ and arbitrary spin $l$, then the potential candidates for $O(x)$ are,
\be\label{fu}
\boxed{
O(x) = (\pa^2)^2 S^{0}(x), \quad  \pa^2 \pa_a S^{1}_{a}(x), \quad  \pa_a S^{3}_{a}(x)} 
\ee
 Later we will argue that we can actually get rid of the operator $(\pa^2)^2 S^{0}(x)$.

\section{Ward Identity for Supertranslation}

The discussion of supertranslation is very similar to the case of the infinite dimensional $U(1)$ symmetry. So let us only mention the changes that are required. 

First of all in the free theory \cite{Banerjee:2018fgd} we can construct the conserved charge $Q_0(f)$ defined as, 
\be
Q_0(f) = \int d\mu(\omega,x) \omega f(x) a^{\dagger}(\omega, x) a(\omega,x)
\ee
where $f(x)$ is an arbitrary function. This is a generalization of the Hamiltonian $H_0$, the generator of time-translation, for which $f(x)=1$. This can be shown to generate supertranslation in the free theory. 

One can check that the commutator of $Q_0(f)$ with the creation operator is given by,
\be
[Q_0(f), a^{\dagger}(\omega,x)] = \omega f(x) a^{\dagger}(\omega,x)
\ee

Similarly under Lorentz transformation $\Lambda$, $Q_0(f)$ transforms as, 
\be
U(\Lambda) Q_0(f) U^{-1}(\Lambda) = Q_0(f'), \qquad  f'(x) = \bigg| \frac{\pa \Lambda^{-1}x}{\pa x}\bigg| ^{-\frac{1}{n}} f(\Lambda^{-1} x)
\ee

The rest of the analysis is identical to that of the $U(1)$ symmetry and so simply quote the final results. 

The conserved charge $Q(f)$ can be written as the sum of a hard and a soft charge, i.e, $Q(f)= Q_H(f)+Q_S(f)$. The hard charge is defined such that,
\be 
\begin{gathered}
Q^{in}_{H}(f) \ket{\alpha, in} = \bigg( \sum_{k\in \alpha} \omega_k f(x_k)\bigg) \ket{\alpha, in} \\
\bra{\beta, out} Q^{out}_{H}(f) = \bra{\beta, out} \bigg( \sum_{k\in \beta} \omega_k f(x_k)\bigg)
\end{gathered}
\ee
Since $Q(f)$ is conserved,
 \be
 Q^{in}_H(f) + Q^{in}_S(f) = Q^{out}_H(f) + Q^{out}_S(f) = Q(f)
  \ee
 This can be written as the Ward-identity \cite{Strominger:2013jfa, He, Strominger:2014pwa, Kapec:2015vwa,Pate:2017fgt, Campiglia:2015kxa,Strominger:2017zoo}, 
\be\label{wst}
\boxed{
 \bra{\beta, out} Q^{out}_S(f)\ket{\alpha, in}  - \bra{\beta, out} Q^{in}_S(f)\ket{\alpha, in} = \bigg( \sum_{i\in\alpha} \omega_i f(x_i) - \sum_{i\in\beta} \omega_i f(x_i) \bigg) \bra{\beta, out}\ket{\alpha, in}}
\ee

It follows from this Ward-identity that the soft charge $Q_S(f)$ transforms under Lorentz transformation as,
\be
\boxed{
U(\Lambda) Q_S(f) U^{-1}(\Lambda) = Q_S(f'), \quad  f'(x) = \bigg| \frac{\pa \Lambda^{-1}x}{\pa x}\bigg| ^{-\frac{1}{n}} f(\Lambda^{-1} x)}
\ee
where we have used that $\Lambda\omega = |\pa x'/\pa x|^{- 1/n} \omega$.

Now we write $Q_S(f)$ as an integral, i.e,
\be
\label{sup_soft_charge}
Q_S(f) = \int d^n x  \, f(x) O(x) 
\ee
where $O(x)$ is a local operator. Then the transformation property of $Q_S(f)$ leads to the following transformation law for $O(x)$ under Lorentz transformation,
\be\boxed{
U(\Lambda) O(x) U(\Lambda)^{-1} = \bigg|\frac{\partial x'}{\partial x}\bigg |^{\frac{\Delta}{n}} O(x') , \quad  x' = \Lambda x, \quad  \Delta = n + 1}
\ee
Therefore, \emph{the Ward-identity \eqref{wst} requires $O(x)$ to be a scalar conformal primary of dimension} $\Delta=n+1$. So this is a \emph{necessary} condition.

\subsection{Potential candidates for $O(x)$ in space-time dimension $D=6$ or $n= D-2 =4$}

Similar to the case of the $U(1)$ symmetry, we now study different possibilities for $O(x)$. For the time being we only consider operators with $\Delta \ge 0$. \textit{A potential candidate for $O(x)$ is a ($\Delta=5,l=0$) primary constructed from} $S^{\Delta}_{a_1a_2....a_l}$.


First let us write down all the $ (\D=5, \, l=0) $ operators starting from $S^{\Delta}_{a_1a_2....a_l}$. They are given by :
\bea
\no \(\D=0 \to \D = 5, \, l=0 \) &:& \hspace{10mm} O(x) = (\pa^2)^2 \pa_a S^{0}_a(x), \hspace{3mm} \pa^2 \pa_a \pa_b \pa_c S^{0}_{abc}(x), \hspace{2mm} \pa_a \pa_b \pa_c \pa_d \pa_e S^{0}_{abcde}(x) \\
\no \(\D=1 \to \D = 5, \, l=0 \) &:& \hspace{10mm} O(x) = (\pa^2)^2 S^{1}(x), \hspace{3mm} \pa^2 \pa_a \pa_b S^{1}_{ab}(x), \hspace{2mm} \pa_a \pa_b \pa_c \pa_d S^{1}_{abcd}(x) \\
\no \(\D=2 \to \D = 5, \, l=0 \) &:& \hspace{10mm} O(x) = \pa^2 \pa_a S^{2}_a(x), \hspace{4mm} \pa_a \pa_b \pa_c  S^{2}_{abc}(x) \\
\no \(\D=3 \to \D = 5, \, l=0 \) &:& \hspace{10mm} O(x) = \pa^2  S^{3}(x), \hspace{8mm} \pa_a \pa_b  S^{3}_{ab}(x) \\
\no \(\D=4 \to \D = 5, \, l=0 \) &:& \hspace{10mm}  O(x) = \pa_a S^{4}_{a}(x)
\eea
Again the requirement that $ O(x) $ must be a $(\Delta=5,l=0)$ primary rules out most of the above operators except the two, $\pa^2 \pa_a \pa_b S^{1}_{ab}(x)$ and $ \pa_a \pa_b  S^{3}_{ab}(x) $. 

The transformation property of the operator $O(x) = \pa_a\pa_b S^3_{ab}(x)$ under infinitesimal SCT \eqref{inf_sct} is
\be
\pa'_a \pa'_b S'^{(3)}_{ab}(x') = (1 + 10 \, \e \cdot x) \, \pa_a  \pa_b \, S^{(3)}_{ab}(x)
\ee
So it is a ($ \D = 5, \, l = 0 $) primary without any constraint. Let us now study the operator $O(x) = \pa^2\pa_a\pa_b S^1_{ab}(x)$. 

\subsection{Operator Decoupling}
Let us rename the operator $S^1_{ab}(x)$ as $h_{ab}(x)$. This can be the leading soft-graviton operator which also transforms like a $(\D=1,l=2)$ primary.

Under an infinitesimal SCT \eqref{inf_sct}, $ h_{ab}(x) $ transforms as
\bea
\no h'_{ab}(x') &=& \left| \f{\pa x'}{\pa x} \right|^{-\f{1}{4}} R_{ac}(x) R_{bd}(x) h_{cd}(x) \\
&=& (1 + 2 \, \e \cdot x) \, h_{ab}(x) + 2 \, \Om_{ac}(x) \, h_{bc}(x) + 2 \, \Om_{bc}(x)h_{ac}(x)
\eea
This gives the transformation law of $O(x) = \pa^2 \pa_a \pa_b h_{ab}(x)$ as,
\be 
O'(x') = \left( 1 + 10 \, \e \cdot x \right) O(x) + \boxed{8 \, \e_a \pa_c\left( \pa_a \pa_b h_{bc}(x) - \pa_b \pa_c h_{ab}(x)\right)}
\ee 
If we define  
\be 
h_a = \pa_b h_{ab}
\ee
and
\be  
F_{ab} = \pa_a h_b - \pa_b h_a
\ee
then we can write,
\be 
O'(x') = \left( 1 + 10 \, \e \cdot x \right) O(x) + \boxed{8 \, \e_a \pa_b F_{ab}(x)}
\ee
So if we want $O(x)$ to be a ($\Delta=5,l=0$) primary then we must set $\pa_a F_{ab}(x) =0$. But $ \pa_a F_{ab} $ is \emph{not} a primary operator and so we cannot set this equal to zero at this stage. 

Under an infinitesimal SCT $\pa_a F_{ab}(x)$ transforms as,
\be\label{F}
\pa'_b F'_{ab}(x') = (1 + 8 \, \e \cdot x) \, \pa_b F_{ab}(x) + 2 \, \Om_{ac} \, \pa_b F_{cb} + \boxed{6 \, \e_b N_{ab}(x)}
\ee
where
\be 
N_{ab} = \pa^2 h_{ab} - \frac{2}{3} \bigg( \pa_a \pa_c h_{cb} + \pa_b \pa_c h_{ca} \bigg) +\frac{1}{3} \delta_{ab} \pa_c\pa_d h_{cd}
\ee

One can now check that \emph{$N_{ab}$ is a $(\Delta=3,l=2)$ primary} and so we can consistently set $N_{ab}(x)=0$, i.e, 
\be\label{EE}
\boxed{
N_{ab} =  \pa^2 h_{ab} - \frac{2}{3} \bigg( \pa_a \pa_c h_{cb} + \pa_b \pa_c h_{ca} \bigg) +\frac{1}{3} \delta_{ab} \pa_c\pa_d h_{cd} = 0}
\ee
This is a (Lorentz) conformally invariant equation. Taking the divergence of this equation we get, 
\be\label{d1}
\boxed{
\pa_a N_{ab} = \frac{1}{3} \pa_a F_{ab} = \frac{1}{3} \pa_a (\pa_a \pa_c h_{cb} - \pa_b \pa_c h_{ca}) =0 }
\ee
So $N_{ab}(x)=0$ implies $\pa_aF_{ab}(x)=0$. \\

Therefore $O(x)= \pa^2 \pa_a\pa_b h_{ab}(x)$ \emph{will be a $(\Delta=5,l=0)$ primary only if $N_{ab}=0$}, i.e, \eqref{EE} holds.  

\subsection{Differential Equation For S-matrix Element}
We proceed exactly as in the case of the $U(1)$.

We denote the S-matrix element with an insertion of $h_{ab}(x)$ in the (incoming) outgoing channel by $\bra{\{\omega_i,x_i, out\}} h_{ab}(x) \ket{\{\omega_j,x_j,in\}}$. For simplicity we also assume that all the incoming and outgoing particles are scalars. 

Let us also define, 
\be
H_{ab}(x, \{\omega_\alpha, x_\alpha\}) = \bra{\{\omega_i,x_i, out\}} h_{ab}(x) \ket{\{\omega_j,x_j,in\}}
\ee
Now we insert the constraint equation \eqref{EE} into an $S$-matrix element and obtain,
\be\boxed{
\bra{\{\omega_i,x_i, out\}} \bigg(\pa^2 h_{ab}(x) - \frac{2}{3} \bigg( \pa_a \pa_c h_{cb}(x) + \pa_b \pa_c h_{ca}(x) \bigg) +\frac{1}{3} \delta_{ab} \pa_c\pa_d h_{cd}(x)\bigg)\ket{\{\omega_j,x_j,in\}} = 0}
\ee 
Since there is no ordering between the $\{x,x_{i,out}, x_{j,in}\}$ coordinates we can pull the derivatives out of the $S$-matrix element. This leads to, 
\be\label{sme2}
\pa^2 H_{ab}(x) - \frac{2}{3} \bigg( \pa_a \pa_c H_{cb}(x) + \pa_b \pa_c H_{ca}(x) \bigg) +\frac{1}{3} \delta_{ab} \pa_c\pa_d H_{cd}(x) = 0
\ee
Here we have suppressed the dependence of $H_{ab}$ on the rest of the variables $\{\omega_\alpha,x_{\alpha}\}$. As far as \eqref{sme2} is concerned these variables are just constants because all the derivatives act on the $x$ coordinate.

If we take the divergence of this equation we get,
\be\label{sme3}
\boxed{
\pa_a f_{ab}(x) = 0} 
\ee
where we have defined, 
\be
f_{ab}(x) = \pa_a H_b - \pa_b H_a = \pa_a \pa_c H_{bc} - \pa_b \pa_c H_{ac}
\ee
Let us note that we could have obtained \eqref{sme3} directly from the operator relation \eqref{d1}. Now in order to solve \eqref{sme2} and \eqref{sme3} we need boundary condition. To obtain the boundary condition we proceed exactly as in the case of $U(1)$ and reproduce only the essential formulas. 

From the Lorentz invariance of S-matrix we can write, 
\be
\bra{\{\omega_i,x_i, out\}} h_{ab}(x) \ket{\{\omega_j,x_j,in\}} = \bra{\{\omega'_i,x'_i, out\}} h'_{ab}(x) \ket{\{\omega'_j,x'_j,in\}}
\ee
where 
\be
\label{fields}
h'_{ab}(x) = \bigg|\frac{\partial x'}{\partial x}\bigg| ^{\frac{1}{4}} R^{-1}_{ac}(x)R^{-1}_{bd}(x) h_{cd}(x') = \bigg|\frac{\partial x'}{\partial x}\bigg| ^{\frac{1}{4}} R_{ca}(x)R_{db}(x) h_{cd}(x')
\ee
 
Now studying the behavior of the $S$-matrix element under inversion $X_a\rightarrow X_a / X^2$ we arrive at the following large $x$ behavior of the $S$-matrix element, 
\be\label{Hf}
\boxed{
\bra{\{\omega_i,x_i, out\}} h_{ab}(x) \ket{\{\omega_j,x_j,in\}} \xrightarrow{\text{$x\rightarrow\infty$}}  \frac{1}{x^2} I_{ac}(x) I_{bd}(x) M_{cd}(\{\omega_\alpha,x_\alpha\}) + O(\frac{1}{x^3})}
\ee
where we have defined 
\be
M_{ab}(\{\omega_\alpha,x_\alpha\}) = \bra{\{\omega'_i,x'_i, out\}} h_{ab}(0) \ket{\{\omega'_j,x'_j,in\}}
\ee
Again $M_{ab}$ is expected to be finite because we have assumed that none of the $(x_{i,out},x_{j,in})$ is at $\infty$. 

Now in terms of $H_{ab}$ the boundary condition \eqref{Hf} can be rewritten as, 
\be\label{bc2}
\boxed{
H_{ab}(x, \{\omega_\alpha, x_\alpha\}) \xrightarrow{\text{$x\rightarrow\infty$}}\frac{1}{x^2} I_{ac}(x) I_{bd}(x) M_{cd}(\{\omega_\alpha,x_\alpha\}) + O(\frac{1}{x^3})}
\ee
As a consistency check, we will show in section \eqref{inv} that the boundary condition \eqref{bc2}, together with Weinberg's soft-graviton theorem, is equivalent to the conservation of energy-momentum. 

Let us now solve the equations \eqref{sme2} and \eqref{sme3}. Equation \eqref{sme3} is the source-free \textit{Euclidean} Maxwell equation on $R^4$ with $H_a$ as the vector potential. Now using the falloff condition on $H_{ab}$ given by \eqref{bc2}, we get $f_{ab}\sim O(\frac{1}{x^4})$ as $x\rightarrow\infty$. So by the same argument as in the case of $U(1)$ (after \eqref{EM}) we get,
\be\label{em}
f_{ab} = \pa_a H_b - \pa_b H_a =0 \Longleftrightarrow  \boxed{H_a = \pa_a \psi(x,\{\omega_\alpha, x_\alpha\}) = \pa_b H_{ab} }
\ee
where $\psi$ is a scalar function. If we substitute this in \eqref{sme2} we get, 
\be\label{poi}
\boxed{
\pa^2 H_{ab} = \frac{4}{3} \bigg( \pa_a \pa_b  - \frac{1}{4} \delta_{ab} \pa^2  \bigg) \psi(x,\{\omega_\alpha,x_\alpha\})}
\ee
This is Poisson's equation for each component of $H_{ab}$. The solution to this equation can be written as, 
\be\label{sol}
H_{ab}(x,\{\omega_\alpha,x_\alpha\}) =\frac{4}{3} \bigg( \pa_a \pa_b  - \frac{1}{4} \delta_{ab} \pa^2  \bigg) f(x,\{\omega_\alpha,x_\alpha\}) + \tilde H_{ab}(x, \{\omega_\alpha, x_\alpha \})
\ee
where, 
\be
\pa^2 f = \psi,  \quad \pa^2 \tilde H_{ab} =0, \quad  \pa_a \tilde H_{ab} = 0
\ee
The last equation, $\pa_a \tilde H_{ab} = 0$, can be obtained by taking divergence of \eqref{sol} and taking into account equation \eqref{em} and the relation $\pa^2 f = \psi$.

Now from the boundary condition \eqref{bc2} we know that $H_{ab}$ goes to zero as $x\rightarrow\infty$. So, if the first term on the R.H.S of \eqref{sol} goes to zero as $x\rightarrow\infty$ then $\tilde H_{ab}$ is must also go to zero as $x\rightarrow\infty$. This will imply that $\tilde H_{ab}$ is identically zero because it is harmonic everywhere in space. 

Now proving that the first term on the R.H.S of \eqref{sol} indeed goes to zero at infinity will require a bit more analysis and so far we have not been able to convince ourselves that this indeed happens. So we keep the general solution for the $S$--matrix in this form, i.e, 
\be
\begin{gathered}
\label{wsg2}
H_{ab}(x, \{\omega_\alpha, x_\alpha\}) =  \\ \boxed{ \bra{\{\omega_i,x_i, out\}} h_{ab}(x) \ket{\{\omega_j,x_j,in\}} 
= \frac{4}{3} \bigg( \pa_a \pa_b  - \frac{1}{4} \delta_{ab} \pa^2  \bigg) f(x,\{\omega_\alpha,x_\alpha\}) + \tilde H_{ab}(x, \{\omega_\alpha, x_\alpha \})}
\end{gathered}
\ee
where $\tilde H_{ab}$ is harmonic everywhere and transverse, i.e, 
\be\label{htg}
\boxed{
\pa^2 \tilde H_{ab}(x, \{\omega_\alpha, x_\alpha\}) = 0 , \quad \pa_a \tilde H_{ab}(x, \{\omega_\alpha, x_\alpha\})  = 0}
\ee

An important point to note is that the term $\tilde H_{ab}$ does \emph{not} contribute to the supertranslation Ward-identity \eqref{wst}. The reason is that the contribution of $h_{ab}$ to the operator $O(x)$ is given by $\pa^2 \pa_a\pa_b h_{ab}$ and so its insertion in a $S$--matrix element is given by,
\be\label{wsf}
\boxed{
\bra{out} O_{in/out}(x) \ket{in} = (\pa^2)^3 f_{in/out}(x, \{\omega_\alpha, x_\alpha\})}
\ee

Now \eqref{wsg2} is also consistent with Weinberg's soft-graviton theorem. In $D=5+1$ Weinberg's soft-graviton theorem can be written as \cite{Kapec:2017gsg,Kapec:2015vwa}, 
\be\label{WHD}
\begin{gathered}
\bra{\{\omega_i,x_i, out\}} h^{out}_{ab}(x) \ket{\{\omega_j,x_j,in\}} \\ 
= \gamma \bigg[ \sum_{i\in out} \omega_i \bigg( \delta_{ab} - 4 \frac{(x-x_i)_a(x-x_i)_b}{(x-x_i)^2} \bigg) - \sum_{j\in in} \omega_j \bigg( \delta_{ab} - 4 \frac{(x-x_j)_a(x-x_j)_b}{(x-x_j)^2} \bigg) \bigg] \\ 
\times \bra{\{\omega_i,x_i, out\}}\ket{\{\omega_j,x_j,in\}}
\end{gathered}
\ee
where $\gamma$ is a numerical constant whose exact value is not necessary for our purpose. Now comparing \eqref{WHD} with \eqref{wsg2} we get,
\be
f(x, \{\omega_\alpha, x_\alpha\}) = \gamma \bigg( \sum_{i\in out} \omega_i (x-x_i)^2 \ln(x-x_i)^2 - \sum_{j \ in} \omega_j (x-x_j)^2 \ln(x-x_j)^2\bigg) \bra{\{\omega_i,x_i, out\}}\ket{\{\omega_j,x_j,in\}} 
 \ee
and
\be
\tilde H_{ab}(x,\{\omega_\alpha,x_\alpha\}) = 0
\ee
So \eqref{wsg2} is in fact a weaker result in the sense that the soft graviton theorem gives $\tilde H_{ab} =0$. It seems that one should be able to prove $\tilde H_{ab}=0$ using \eqref{htg} and the boundary conditions. We hope to return to this in future. 

\subsection{Other Operators In The List}

\bea
\no \(\D=0 \to \D = 5, \, l=0 \) &:& \hspace{10mm} O(x) = (\pa^2)^2 \pa_a S^{0}_a(x), \hspace{3mm} \pa^2 \pa_a \pa_b \pa_c S^{0}_{abc}(x), \hspace{2mm} \pa_a \pa_b \pa_c \pa_d \pa_e S^{0}_{abcde}(x) \\
\no \(\D=1 \to \D = 5, \, l=0 \) &:& \hspace{10mm} O(x) = (\pa^2)^2 S^{1}(x), \hspace{3mm} \pa^2 \pa_a \pa_b S^{1}_{ab}(x), \hspace{2mm} \pa_a \pa_b \pa_c \pa_d S^{1}_{abcd}(x) \\
\no \(\D=2 \to \D = 5, \, l=0 \) &:& \hspace{10mm} O(x) = \pa^2 \pa_a S^{2}_a(x), \hspace{4mm} \pa_a \pa_b \pa_c  S^{2}_{abc}(x) \\
\no \(\D=3 \to \D = 5, \, l=0 \) &:& \hspace{10mm} O(x) = \pa^2  S^{3}(x), \hspace{8mm} \pa_a \pa_b  S^{3}_{ab}(x) \\
\no \(\D=4 \to \D = 5, \, l=0 \) &:& \hspace{10mm}  O(x) = \pa_a S^{4}_{a}(x)
\eea

We have already considered the operator $\pa^2 \pa_a \pa_b S^1_{ab}(x)$ and $\pa_a\pa_b S^3_{ab}(x)$. The rest of the operators in the list can be studied in the same way. Let us, for example, consider the operator $O(x) = (\pa^2)^2 S^1(x)$. Under infinitesimal SCT \eqref{inf_sct} the operator transforms as, 
\be
(\pa'^2)^2 S'^{1}(x') =(1 + 10 \, \e \cdot x) \, (\pa^2)^2 \, S^{1}(x) + \boxed{8 \, \e_a \pa_a \pa^2 \, S^{1}(x)} 
\ee
So if we want this operator to transform as a primary then we have to set the extra piece $\pa_a\pa^2 S^1(x)$ to zero. But this cannot be done because $\pa_a\pa^2 S^1(x)$ is not a primary operator.  Under an infinitesimal SCT \eqref{inf_sct} it transforms as, 
\be
\pa'_a \pa'^2 S'^{1}(x') = (1+ 8 \, \e \cdot x) \, \pa_a  \pa^2 \, S^{1}(x) + 2\, \Om_{ab} \pa_b \pa^2 \, S^{1}(x) + \boxed{6\, \e_a \pa^2 \, S^{1}(x)}
\ee
So we have to set $\pa^2 S^1(x)$ to zero and this is possible because it is a primary operator, i.e, under infinitesimal SCT \eqref{inf_sct}
\be
\pa'^2 S'^{1}(x') = (1 + 6 \, \e \cdot x) \, \pa^2 \, S^{1}(x)
\ee
Now we can impose the constraint $ \pa^2 \, S^{1}(x) = 0 $ consistently but this makes the operator $O(x)= (\pa^2)^2 S^1(x)$ identically zero. 

Therefore the potential candidates for $O(x)$ in the $\D\ge 0$ range are $O(x) = \pa^2 \pa_a \pa_b S^{1}_{ab}(x)$ and $\pa_a \pa_b  S^{3}_{ab}(x)$.

Actually there is one more candidate for $O(x)$ which is constructed out of a $(\D=-1,l=0)$ primary $S^{-1}(x)$, given by $O(x) = (\pa^2)^3 S^{-1}(x)$. One can check that this indeed is a $(\D=5,l=0)$ primary without any constraint on $S^{-1}(x)$. We have not mentioned this in this section because it has scaling dimension $\D=-1$. 

Therefore the complete list of possibilities for $O(x)$ is,
\be\boxed{
O(x) = \pa^2 \pa_a \pa_b S^{1}_{ab}(x), \quad  (\pa^2)^3 S^{-1}(x), \quad  \pa_a \pa_b  S^{3}_{ab}(x)} 
\ee

Later we will argue that we can actually get rid of the operator $(\pa^2)^3 S^{-1}(x)$.

\section{$\D<0$ primaries and more general possibilities for $O(x)$}
\label{null}
So far, in both the cases of $U(1)$ and supertranslation, we have studied only primary operators $S^{\D}_{a_1a_2..a_l}(x)$ with $\D\ge 0$. Now we want to relax the constraint on $\D$. In order to do this we need to systematise our observations in the last sections because for $\D < 0$ there are an infinite number of operators to check. Since the discussion for the $U(1)$ and supertranslation are structurally identical, as far the representation of the conformal group is concerned, let us focus on $U(1)$ : 


In space-time dimensions $D=5+1$ or $n=4$, the Ward-identity \eqref{u1} requires the existence of a $(\D=4,l=0)$ \emph{primary} $O(x)$ which we assume to be a descendant of the primaries $S^{\D}_{a_1a_2...a_l}(x)$. If we start with a primary $S^{\D}_{a_1a_2...a_l}(x)$ then we can construct a unique $(\Delta=4,l=0)$ descendant $O(x)$ given by,
\be\label{pc}
O(x) = (\pa^2)^{\frac{4-\D-l}{2}} \pa_{a_1}....\pa_{a_l} S^{\D}_{a_1a_2...a_l}(x)
\ee 
This requires $\frac{4-\D-l}{2}$ to be a non-negative integer. Therefore not all primaries $S^{\D}_{a_1a_2...a_l}(x)$ can contribute to $O(x)$. Now we impose the restriction that $O(x)$ should also be a \emph{primary} operator. This requires $S^{\D}_{a_1a_2...a_l}(x)$ to have a $(\D=4,l=0)$ \emph{primary descendant}. This is a non-trivial constraint. To solve this we study how $O(x)$ in \eqref{pc} transforms under infinitesimal special conformal transformations. At the first step this generates the operator $K_{b_1}O$ where $K_{b_1}$ is a generator of SCT. If $O(x)$ is a primary then $K_{b_1}O=0$ and the process stops. If not then we repeat it. At the next step we generate the operator $K_{b_2}K_{b_1}O$. If $K_{b_1}O$ is a primary then $K_{b_2}K_{b_1}O = 0$ and the process stops. Since $K_{b_1}O$ is a primary we can consistently set $K_{b_1}O=0$ and this turns  $O(x)$ into a primary. The equation $K_{b_1}O=0$ becomes a linear conformally invariant differential equation which is satisfied by the corresponding primary $S^{\D}_{a_1a_2...a_l}(x)$. Now a crucial point is that the constraint $K_b O=0$ can make the operator $O$ vanish unless the dimension and the spin of the primary $S^{\D}_{a_1a_2...a_l}(x)$ take certain specific values. For example, we have seen that  when $O(x)= \pa^2\pa_aS^1_a(x)$, the constraint equation is $K_b O = \pa_a (\pa_a S^1_b - \pa_b S^1_a)= \pa_a F_{ab}=0$. Now $\pa_a F_{ab}$ is a $(\D=3,l=1)$ primary and it has a unique scalar descendant of dimension $4$, given by $\pa_b\pa_aF_{ab}$, which is identically zero because of the antisymmetry of $F_{ab}$. Now, if this were not zero, then this would have been proportional to the operator $O(x)$ and so the constraint $\pa_aF_{ab}=0$ would make $O(x)$ zero, which is a trivial solution. In fact this is exactly the case for the primaries $S^{\D}_{a_1a_2...a_l}(x)$ with $\D < 0$. 


Now suppose that instead of the second step, the process stops after $(p+1)$ steps, i.e, $K_{b_{p+1}}\big(K_{b_p}K_{b_{p-1}}...K_{b_1}O\big)=0$. So the primary operator $K_{b_p}K_{b_{p-1}}...K_{b_1}O$ is obtained at the $p$-th step. Since $O(x)$ is a scalar and $K_{a}$'s commute, the primary operator $K_{b_p}K_{b_{p-1}}...K_{b_1}O$ must be a symmetric traceless tensor of some rank $p'\le p$. \footnote{It may happen that $K_{b_p}K_{b_{p-1}}...K_{b_1}O$ is a sum of product of Kronecker deltas and primaries of weight $(4-p)$ and rank $p'\le p$. For example consider the operator $O(x)= (\pa^2)^2\pa_aS^{-1}_a$. The primary operator that one gets in this case is of the form, $K_{b_4}...K_{b_1}O = \delta_{(b_1b_2}S^{0}_{b_3b_4)}$ where $S^{0}_{ab}=S^{0}_{ba} = \pa_a S^{-1}_b + \pa_b S^{-1}_a - \frac{1}{2} \delta_{ab} \pa\cdot S^{-1}$. $S^{0}_{ab}$ is a $(\D=0,l=2)$ primary which, according to our notation, should be denoted by $O^0_{b_1b_2}$.} Let us denote this $(\D=4-p,l=p'\le p)$ primary by $O^{\delta}_{b_1b_2...b_{p'}}$ where $\delta=4-p$ is the dimension of the operator. Now the important point is that $O^{\delta}_{b_1b_2...b_{p'}}$ is also a descendant of $S^{\D}_{a_1a_2...a_l}$. Therefore $O^{\delta}_{b_1b_2...b_{p'}}$ is a \emph{primary descendant} of $S^{\D}_{a_1a_2...a_l}$, which transforms in the \emph{symmetric traceless tensor representation}. Therefore, in order to find out if $O^{\delta}_{b_1b_2...b_{p'}}$ exists we need to study the \emph{primary descendants} of $S^{\D}_{a_1a_2...a_l}$ \emph{which transform in the symmetric traceless tensor representation}. This problem was completely solved in a $d$ dimensional Euclidean CFT for all values of $\D$ and $l$ in \cite{Penedones:2015aga}.

According to \cite{Penedones:2015aga} if we start with a $(\D,l)$ primary $S^{\D}_{a_1a_2...a_l}$ in an Euclidean $CFT_d$ -- $d=4$ in this paper -- then there are three types of primary descendants that one can obtain depending on the values of $\D$ and $l$. They are given as follows : 

1) Type--1 : These are $(\D + n,n+l)$ primary descendants $\tilde S^{\D+n}_{a_1a_2...a_{l+n}}$ which occur when 
\be
\D = 1-l-n , \quad n=1,2,3,......
\ee

2) Type--2 : These are $(\D + n,l-n)$ primary descendants $\tilde S^{\D+n}_{a_1a_2...a_{l-n}}$ which occur when 
\be
\D = l+d-1-n = l-n +3 , \quad n=1,2,3,......,l
\ee

3) Type--3 : These are $(\D + 2n,l)$ primary descendants $\tilde S^{\D+2n}_{a_1a_2...a_l}$ which occur when 
\be
\D = \frac{d}{2} -n = 2-n , \quad n=1,2,3,......
\ee
Let us now study the primary descendants when $\D$ is an integer $< 0$. Since the operators $O^{\delta}_{b_1b_2...b_{p'}}$ we are looking for are also obtained by SCT of the $\D=4$ scalar operator $O(x)$, we need to focus on primary descendants of $S^{\D}_{a_1a_2...a_l}$ with dimension $\delta < 4$. We now discuss different possibilities : \\

3) For primary descendants of type--3 we have $\D = 2-n \rightarrow n= 2-\D$. Therefore the dimension of the primary descendant $\tilde S^{\D+2n}_{a_1a_2...a_l}$ is $\D + 2n = 4-\D > 4$ for $\D<0$. Since the dimension of $O^{\delta}_{b_1b_2...b_{p'}}$ this case is ruled out. \\


2) For primary descendants of type--2 we have $\D = l-n+3 \rightarrow l-n= \D-3 < 0$ because $\D <0$. But a primary descendants of type--2 exists only if $n\le l$, i.e, $l-n \ge 0$. Therefore no primary descendants of type--2 exists for $\D <0$. \\

1) So for $\D<0$ we need to study primary descendants of type--1. Now $O^{\delta}_{b_1b_2...b_{p'}}$ could be a primary descendant of type--1 or it could be a primary descendant of type--2 or type--3 of another primary descendant of $S^{\D}_{a_1a_2...a_l}$ of type --1with dimension $<\delta$. But such cases are not relevant for our purpose. Let us now consider below different possibilities : \\

1a) First consider the $(\D=-1,l=1)$ primary $S^{-1}_a(x)$. This gives a potential candidate for $O(x)$ given by $(\pa^2)^2\pa_aS^{-1}_a(x)$. Now $S^{-1}_a(x)$ has a $(\D=0,l=2)$ type--1 primary descendant $S^{0}_{ab}$ given by,
\be
S^{0}_{ab}(x) = \pa_a S^{-1}_b(x) + \pa_b S^{-1}_a(x) - \frac{1}{2} \delta_{ab} \pa\cdot S^{-1}(x)
\ee
Now we have to look for primary descendants of $S^{0}_{ab}$ of scaling dimension $< 4$. One can easily check that there are no such primary descendants of $S^0_{ab}(x)$ of dimension $< 4$. So if $O^{\delta}_{b_1b_2...b_{p'}}$ exists then it can only be $S^{0}_{ab}(x)$. If it does not, then we reach $S^{-1}_{a}(x)$ by repeated SCT of $(\pa^2)^2\pa_aS^{-1}_a(x)$ and there is no way by which we can turn $(\pa^2)^2\pa_aS^{-1}_a(x)$ into a primary. Now suppose this is not the case. Then by applying four SCTs to the operator $O(x)=(\pa^2)^2\pa_aS^{-1}_a(x)$ we will reach the primary descendant $S^0_{ab}$ which we can consistently set to zero. But this is a trivial solution to our problem for the following reason. Consider the operator $\pa^2 \pa_a\pa_b S^0_{ab}(x)$. It is easy to see that,
\be
O(x) = (\pa^2)^2\pa_aS^{-1}_a(x) = \frac{2}{3} \pa^2 \pa_a\pa_b S^0_{ab}(x)
\ee 
Therefore if we set $S^0_{ab}$ to zero then the operator $O(x)=(\pa^2)^2\pa_aS^{-1}_a(x) =0$, which is the trivial solution.  

So the operator $S^{-1}_{a}(x)$ cannot contribute to the soft charge $Q_S(f)$ for the $U(1)$ symmetry.

We have also checked these conclusions by brute force calculation, i.e, by directly applying SCT to the operator $O(x)=(\pa^2)^2\pa_aS^{-1}_a(x)$. In fact after applying four SCT we get the $(\D=0,l=2)$ primary $\delta_{(ab}S^0_{cd)}(x)$.

The same thing happens in the case of the $(\D=-3,l=1)$ primary $S^{-3}_a(x)$. Please see the Appendix for details. Let us now study the $(\D \le -5,l=1)$ primaries. \\

1b) For $(\D\le -5,l=1)$ there are not even any primary descendants of type -- 1 which is relevant in our case. By starting with with a $S^{-5}_{a}(x)$ we generate a $(\D=0,l=6)$ type -- 1 primary descendant. Now since $O(x)$ has dimension $4$, we can generate at most a spin -- 4 primary descendant with dimension $0$, by applying four SCT. So this is ruled out. One can easily check that the same argument holds for any $\D\le -5$. \\

1c) In the same way one can check that primary operators $S^{\D}_{a_1a_2...a_l}$ with $\D < 0$ and $l \neq 1$ cannot contribute to the soft charge of $U(1)$. \\

Therefore in the case of the $U(1)$ symmetry the only possible candidates are: 
\be\label{fu2}
\boxed{
O(x) = (\pa^2)^2 S^{0}(x), \quad  \pa^2 \pa_a S^{1}_{a}(x), \quad  \pa_a S^{3}_{a}(x)} 
\ee

For the case of supertranslations the relevant negative dimension primaries are $(\D=-3,l=2)$ and $(\D=-5,l=2)$. These two can also be ruled out and the details are given in the Appendix. 

So, the only possible candidates for $O(x)$ in case of supertranslation are: 
\be\label{su2}
\boxed{
O(x) = \pa^2 \pa_a \pa_b S^{1}_{ab}(x), \quad  (\pa^2)^3 S^{-1}(x), \quad  \pa_a \pa_b  S^{3}_{ab}(x)} 
\ee

\section{"Gauge--structure" in differential equations and "redundancy" of scalar primaries}
The differential equations \eqref{smc11} and \eqref{EE}, obtained by setting the primary descendants of $S^{1}_a(x)=S_a(x)$ and $S^1_{ab}(x)=h_{ab}(x)$ to zero, are both equations of motion of Euclidean "gauge -- theories". Both have scalar "gauge-invariance" \cite{Erdmenger:1997wy,Dolan:2001ih,Beccaria:2015vaa} , i.e, the gauge transformation function is a scalar in both cases. 

\subsection{The case of $U(1)$}
The operator $S_a(x)$ satisfies \eqref{smc11}, 
\be\label{ME}
\pa_a F_{ab} = \pa_a (\pa_a S_b - \pa_b S_a) =0
\ee
This is Euclidean Maxwell equation with $S_a(x)$ as the "gauge" potential. The word gauge is within quotation marks because there is \emph{no} obvious redundancy in our description because $S_a(x)$ is supposed to be built out of the creation/annihilation operators of \emph{physical} photons of helicity $a$. We suspect that this is somehow related to the large $U(1)$ gauge transformations at null-infinity. It will be very interesting and important to clarify this connection. \\

 For our purpose, this "redundancy" has an interesting application. Let us recall that we obtained \eqref{ME} by demanding that the operator $O(x)= \pa^2\pa_aS_a(x)$ be a $(\D=4,l=0)$ primary. Now \eqref{ME} is invariant under the familiar substitution $S_a(x)\rightarrow \tilde S_a(x) = S_{a}(x) + \pa_a \phi(x)$. This is a valid substitution only if $\pa_a\phi(x)$ also transforms like a $(\D=1,l=1)$ primary. It is easy to check that this is possible only if $\phi(x)$ is a $(\D=0,l=0)$ primary. So we can see that if $O(x)=\pa^2\pa_aS_a(x)$ is a $(\D=4,l=0)$ primary then $\tilde O(x) = \pa^2\pa_a\tilde S_a(x)$ is also a $(\D=4,l=0)$ primary if we choose $\phi(x)$ to be $(\D=0,l=0)$ primary. 

Now, we have seen that there are three potential candidates for $O(x)$ given by $ (\pa^2)^2 S^{0}(x)$, $\pa^2 \pa_a S_{a}(x)$ and $ \pa_a S^{3}_{a}(x)$. So generically it is true that,
\be
O(x) \supseteq  \pa^2 \pa_a S_{a}(x) + \alpha (\pa^2)^2 S^{0}(x) + \beta \pa_a S^{3}_{a}(x)
\ee
where $\alpha$ and $\beta$ are non-zero numbers. Let us note that $S^0(x)$ is a $(\D=0,l=0)$ primary which can play the role of $\phi(x)$. So we \emph{redefine} our $S_a(x)$ as, 
\be\boxed{
S_a(x) \rightarrow \tilde S_a(x) = S_a(x) + \alpha \pa_a S^0(x)}
\ee
With this redefinition we can write,
\be
O(x) =  \pa^2 \pa_a \tilde S_{a}(x) + \beta \pa_a S^{3}_{a}(x)
\ee
Since $\tilde S_a(x)$ is a $(\D=1,l=1)$ primary which also satisfies \eqref{ME}, $\pa^2 \pa_a \tilde S_{a}(x)$ is again a $(\D=4,l=0)$ primary. 

Therefore we can conclude that \emph{\underline{there exists a basis} of primary operators $S^{\D}_{a_1a_2.....a_l}$ in which the the most general form of the operator $O(x)$ is given by},
\be\label{pf}
\boxed{
O(x) =  \pa^2 \pa_a S^1_{a}(x) + \beta \pa_a S^{3}_{a}(x)}
\ee
This is a pure spin-$1$ contribution. This is possible only because the constraint equation \eqref{ME}, which is Maxwell's equation in this case, has the right kind of redundancy. 

\subsection{The case of supertranslation}
Let us now study the equation \eqref{EE} satisfied by $h_{ab}(x)$, 
\be\label{EE2}
\pa^2 h_{ab} - \frac{2}{3} \bigg( \pa_a \pa_c h_{cb} + \pa_b \pa_c h_{ca} \bigg) +\frac{1}{3} \delta_{ab} \pa_c\pa_d h_{cd} = 0
\ee
This is also the equation of motion of an Euclidean gauge--theory in which the gauge--transformation is given by,
\be\label{sg}
\boxed{
h_{ab}(x) \rightarrow \tilde h_{ab}(x) = h_{ab}(x) + \bigg( \pa_a\pa_b - \frac{1}{4} \delta_{ab} \pa^2 \bigg) \phi(x)}
\ee
This is known as "scalar gauge invariance" \cite{Erdmenger:1997wy,Dolan:2001ih,Beccaria:2015vaa} because the gauge transformation parameter is an arbitrary scalar $\phi(x)$. In fact, \eqref{EE2} is the \emph{equation of motion of a conformal symmetric tensor} $h_{ab}$ \cite{Beccaria:2015vaa} with gauge invariance given by \eqref{sg}. So in some sense \emph{$h_{ab}(x)$ can be thought of as a conformal symmetric tensor living on the celestial sphere}. Again, the "gauge-transformation" has no obvious origin except that this is somehow related to supertranslation at null-infinity which is also determined by a single function \cite{Kapec:2015vwa,Pate:2017fgt}. Clarification of this connection seems to be of utmost importance. Let us now get rid of the scalar contribution to $O(x)$ using this "redundancy".  \\
  
Since $h_{ab}(x)$ is a $(\D=1,l=2)$ primary, the substitution \eqref{sg} is valid only if $\tilde h_{ab}(x)$ is also a $(\D=1,l=2)$ primary. One can easily check that this is possible only if the scalar $\phi(x)$ is a $(\D=-1,l=0)$ primary. If this true then \emph{both the operators, $O(x)=\pa^2 \pa_a\pa_b h_{ab}(x)$ and $\tilde O(x)=\pa^2 \pa_a\pa_b \tilde h_{ab}(x)$, are $(\D=5,l=0)$ primary}. It is clear that the invariance of the constraint equation \eqref{EE2} under the substitution \eqref{sg} is crucial for this. 
Now let us recall that the operator $O(x)$, in the case of supertranslation can be written as, 
\be
O(x)=\pa^2 \pa_a \pa_b h_{ab}(x) + \alpha (\pa^2)^3 S^{-1}(x) + \beta\pa_a \pa_b  S^{3}_{ab}(x)
\ee
where $\alpha$ and $\beta$ are non-zero numbers. Let us also note that here $S^{-1}(x)$ is a $(\D=-1,l=0)$ primary which can play the role of $\phi(x)$. So we \emph{redefine} $h_{ab}(x)$ as,
\be\boxed{
h_{ab}(x) \rightarrow \tilde h_{ab}(x) = h_{ab}(x) + \alpha\frac{4}{3} \bigg( \pa_a\pa_b - \frac{1}{4} \delta_{ab} \pa^2 \bigg) S^{-1}(x)}
\ee
In terms of the redefined operator we can write, 
\be
O(x)=\pa^2 \pa_a \pa_b \tilde h_{ab}(x) + \beta\pa_a \pa_b  S^{3}_{ab}(x)
\ee
Therefore we can conclude that \emph{\underline{there exists a basis} of primary operators $S^{\D}_{a_1a_2.....a_l}$ in which the the most general form of the operator $O(x)$ is given as},
\be\label{gf}
\boxed{
O(x)=\pa^2 \pa_a \pa_b S^1_{ab}(x) + \beta\pa_a \pa_b  S^{3}_{ab}(x)}
\ee
This is a pure spin-$2$ contribution. 

Now let us mention that the "gauge-structure" may somehow be related to the angle-dependent large $U(1)$ gauge transformations 
and supertranslation at null-infinity. In both cases the gauge transformation parameter is a scalar function defined on a space-like cross-section of the null-infinity. We leave further study of this to future work. 

\section{Can the Ward-identity be solved?}
For concreteness let us focus on the $U(1)$ symmetry. We have seen that if we have some energetic charged particles and a conformal sector with primary operators $S^{\D}_{a_1a_2...a_l}$ then the infinite dimensional $U(1)$ symmetry and Lorentz invariance of the $S$--matrix dictates the form of the soft-charge to be \eqref{pf}, 
\be
Q^{out}_S(f) = \int d^4x \ f(x) O^{out}(x) = \int d^4x f(x) \bigg( \gamma\pa_a S^{3}_{a,out}(x) + \delta \pa^2\pa_aS^{1}_{a,out}(x) \bigg)
\ee
\be
Q^{in}_S(f) = \int d^4 x \ f(x) O^{in}(x) = \int d^4x f(x) \bigg( \gamma'\pa_a S^{3}_{a,in}(x) + \delta' \pa^2\pa_aS^{1}_{a,in}(x) \bigg)
\ee
where $(\gamma,\delta)$ and $(\gamma',\delta')$ are real numbers. Now let us write the Ward-identity \eqref{u1} in the unintegrated form,
\be\label{fde1}
\bra{\beta, out} O^{out}(x)\ket{\alpha, in}  - \bra{\beta, out} O^{in}(x)\ket{\alpha, in} = \bigg( \sum_{i\in\alpha} q_i \delta^4(x-x_i) - \sum_{j\in\beta} q_j \delta^4(x-x_j) \bigg) \bra{\beta, out}\ket{\alpha, in}
\ee
Now this is a differential equation for the $S$-matrix elements with the insertion of soft-operators. This may or may not be solvable depending on the structure of $O(x)$. For example, if we take $O(x)=\pa_aS^3_a(x)$ then there is no way to solve this equation because there is one differential equation and four (or eight) unknown functions. 

The \emph{simplest} theory corresponds to the choice $O(x)= \pa^2\pa_aS_a(x)$. In this case we know, from the decoupling of primary descendants, that \eqref{scu} holds, i.e, 
\be\boxed{
\bra{\beta, out} S_{a,out}(x)\ket{\alpha, in} = \pa_a \Lambda_{out}(x)}  \quad \boxed{
\bra{\beta, out} S_{a,in}(x)\ket{\alpha, in} = \pa_a \Lambda_{in}(x)}
\ee
Here we have omitted the dependence of $\Lambda_{in/out}$ on the coordinates of in and out particles. Now Substituting these in the Ward-identity \eqref{fde1} we get, 
\be\label{wsg}
(\pa^2)^2 \Lambda(x) = \bigg( \sum_{i\in\alpha} q_i \delta^4(x-x_i) - \sum_{j\in\beta} q_j \delta^4(x-x_j) \bigg) \bra{\beta, out}\ket{\alpha, in}
\ee
where we have defined,
\be
\Lambda = \delta\Lambda_{out} - \delta'\Lambda_{in}
\ee
This equation can be easily solved subject to the boundary condition \eqref{moff}, i.e, 
\be
\pa_a\Lambda(x) \xrightarrow{\text{$x\rightarrow\infty$}}  \frac{1}{x^2} I_{ab}(x) M_b + O(\frac{1}{x^3})
\ee
where $M_a$ is some constant vector dependent only on the coordinates of the hard particles. 

Now one can solve this equation and check that $\pa_a\Lambda$ is given by Weinberg's soft-photon theorem, upto normalisation. 

Now $\pa_a\Lambda$ is given by,
\be
\pa_a\Lambda = \delta \pa_a\Lambda_{out} - \delta'\pa_a\Lambda_{in} = \delta \bra{\beta, out} S_{a,out}(x)\ket{\alpha, in} - \delta' \bra{\beta, out} S_{a,in}(x)\ket{\alpha, in}
\ee
and so is a linear combination of $S$--matrix elements with an outgoing and an incoming soft-photon, respectively. These two amplitudes should be related by crossing but we are unable to determine the precise relation between them at this point. But, we believe that this can be done, and solution to the Ward-identity, with $O(x)= \pa^2\pa_aS_a(x)$, is given by Weinberg's soft-photon theorem (upto normalization). It will be interesting to complete this proof. 

Without going into detail let us simply mention that the same conclusion can be reached for the case of supertranslation in essentially the same way. 

\section{The case for space-time dimension, $D = 3+1$}
In four space-time dimensions things are somewhat different, although the final answers are the same. In this case the Lorentz group $SO(3,1)$ acts as the (global) conformal group on $R^2$ on which the soft-operators live. Let us describe the case of $U(1)$ and supertranslation separately. 

\subsection{$U(1)$}
In this case the operator $O(x)$ should be primary operator of dimension $\D=n=D-2=2$. Now starting from the $(\D=1,l=1)$ primary $S^1_a(x)=S_a(x)$ we can construct \emph{two different scalar primaries} of dimension $2$ given by, 
\be
O_1(x) = \pa_a S_a(x) , \quad O_2(x) = \epsilon_{ab} \pa_a S_b(x)
\ee
where $\epsilon_{ab}$ is the antisymmetric tensor with $\epsilon_{12}=1$. Now we can easily see the difference with the six dimensional case. In $D=5+1$ we had a single scalar operator $\pa^2\pa_aS_a(x)$ which was not a primary to start with and we had to impose the decoupling condition of the primary descendant $\pa_a (\pa_a S_b(x) - \pa_b S_a(x))$ to make it primary. So in $D=5+1$ the equation $\pa_a (\pa_a S_b(x) - \pa_b S_a(x))=0$ was guaranteed by Lorentz invariance of the theory. 

Now in $D=4$, in keeping with the higher dimensions we can set $O_2$ to zero , i.e, 
\be\boxed{
O_2(x) = \epsilon_{ab} \pa_a S_b(x) =0} 
\ee
This immediately gives, 
\be
\bra{\beta, out} S^1_{a,out}(x)\ket{\alpha, in} = \pa_a \Lambda_{out}(x)
\ee
\be
\bra{\beta, out} S^1_{a,in}(x)\ket{\alpha, in} = \pa_a \Lambda_{in}(x)
\ee
just like in $D=5+1$. 

The crucial difference with higher dimensions is that decoupling of $O_2$ is \emph{not} required by conformal invariance. But we can still decouple $O_2$ by demanding that the Ward-identity be solvable. 

\subsection{Supertranslation}
In this case $O(x)$ is scalar primary of dimension $\D=n+1=3$ constructed as a descendant of the $(\D=1,l=2)$ primary $h_{ab}(x)$. One can construct three scalar operators with dimension $\D=3$, all of which are \emph{primaries}. They are given by, 
\be
\boxed{
O_1(x) = \pa_a\pa_b h_{ab}(x) , \quad O_2(x) = \epsilon_{ab} \pa_a \pa_c h_{cb}(x)} , \quad O_3(x) = \epsilon_{ab}\epsilon_{cd}\pa_a\pa_c h_{bd}(x)
\ee
Now one can check using the tracelessness of $h_{ab}(x)$ that $O_1(x)\propto O_3(x)$. So there are two independent $(\D=3,l=0)$ primaries given by $O_1(x)$ and $O_2(x)$. Now, in keeping with the high dimensional case, let us set $O_2(x)$ to zero, i.e, 
\be\label{ck}
\boxed{
O_2(x) = \epsilon_{ab} \pa_a \pa_c h_{cb}(x) =0} 
\ee
It is reassuring that, just like in $D=5+1$, \eqref{ck} is also invariant under the "gauge transformation",
\be\label{p2d}
\boxed{
h_{ab}(x) \rightarrow h'_{ab}(x) = h_{ab}(x) + \bigg( \pa_a\pa_b - \frac{1}{2} \delta_{ab} \pa^2\bigg) \phi (x)}
\ee

Now \eqref{ck} can be converted into an equation for the $S$-matrix element with an insertion of $h_{ab}(x)$. Let us define,
\be
H_{ab}(x)= \bra{out}h_{ab}(x)\ket{in}
\ee
Here $h_{ab}$ could be either an incoming or an outgoing soft-graviton. We have also omitted the dependence of $H_{ab}$ on the coordinates of the hard particles. Therefore we can write,
\be\label{cks}
\epsilon_{ab} \pa_a \pa_c H_{cb}(x) =0
\ee
If we define $H_a = \pa_b H_{ab}$, then we can rewrite \eqref{cks} as, 
\be
\epsilon_{ab} \pa_a H_b(x) =0 \Rightarrow \boxed{H_a(x) = \pa_b H_{ab}(x) = \pa_a \Lambda(x)} 
\ee
It is easy to solve this equation in complex coordinates on the plane. If take into account the fact that $H_{ab}$ is traceless, i.e, $H_{z \bar z}=0$, then we can write,
\be\label{2}
\pa_{\bar z} H_{zz} = \pa_z \Lambda , \quad \pa_{z} H_{\bar z \bar z} = \pa_{\bar z} \Lambda
\ee
Let us consider the first equation and multiply both sides by $\pa_z$. Then this leads to, 
\be
\pa^2 H_{zz} = \pa_z \pa_z \Lambda
\ee
The solution of this equation can be written as, 
\be\label{1}
H_{zz} = \pa_z \pa_z \psi + \tilde H_{zz}
\ee
where $\pa^2\psi = \Lambda$ and $\pa^2 \tilde H_{zz}=0$. In fact, $\tilde H_{zz}$ is holomorphic. To see this, we act on \eqref{1} with $\pa_{\bar z}$ and use \eqref{2} and $\pa^2\psi=\Lambda$, to get
\be
\pa_{\bar z} H_{zz} = \pa_{\bar z} \pa_z \pa_z \psi + \pa_{\bar z} \tilde H_{zz} \Rightarrow \pa_{\bar z} \tilde H_{zz} =0 
\ee
The second equation can also be solved in the same way and is determined in terms of the \emph{same} function $\psi$. 

Therefore the most general solution for the $S$--matrix element $H_{zz}$ can be written as, 
\be\boxed{
H^{in}_{zz} = \bra{out}h^{in}_{zz}\ket{in} = \pa_z \pa_z \psi^{in} (z,\bar z)} + \tilde H_{zz} (z)
\ee
\be\boxed{
H^{in}_{\bar z\bar z} = \bra{out}h^{in}_{\bar z\bar z}\ket{in} = \pa_{\bar z} \pa_{\bar z} \psi^{in} (z,\bar z)} + {\tilde H'}_{\bar z\bar z} (\bar z)
\ee
\be\boxed{
H^{out}_{zz} = \bra{out}h^{out}_{zz}\ket{in} = \pa_z \pa_z \psi^{out} (z,\bar z)} + \tilde H_{zz} (z)
\ee
\be\boxed{
H^{out}_{\bar z\bar z} = \bra{out}h^{out}_{\bar z\bar z}\ket{in} = \pa_{\bar z} \pa_{\bar z} \psi^{out} (z,\bar z)} + {\tilde H'}_{\bar z\bar z} (\bar z)
\ee
Here we have restored the in/out index.

Now, since $(\tilde H')$ $\tilde H$ is (anti-) holomorphic, it does not contribute to the Ward-identity because it cancels in the matrix element of the charge or the operator $O(x)$. To see this, $O(x)$ can be written as $O(z,\bar z) = \pa_{\bar z}\pa_{\bar z} h_{zz} + \pa_z \pa_z h_{\bar z\bar z}$ and so,
\be
\bra{out} O^{in}(z,\bar z)\ket{in} =  (\pa^2)^2 \psi^{in}(z,\bar z) + (\pa^2)^2 \psi^{in}(z,\bar z)
\ee
\be
\bra{out} O^{out}(z,\bar z)\ket{in} =  (\pa^2)^2 \psi^{out}(z,\bar z) + (\pa^2)^2 \psi^{out}(z,\bar z)
\ee
\\

So far we have only talked about the primary descendants or null states under the global conformal group $SL(2,\mathbb C)$. But what about the Virasoro algebra? In fact in $D=3+1$ it has been conjectured that the $SL(2,\mathbb C)$ conformal group gets enhanced to the full Virasoro algebra or the algebra of super-rotations \cite{Barnich:2011ct, Kapec:2016jld}. Decoupling of Virasoro null-states has a far richer story. So it is natural to wonder what happens in $D=3+1$ if instead of $SL(2,\mathbb C)$ we let the Virasoro null-states to decouple. May be this is more natural. In fact we have seen that in $D=3+1$, Lorentz invariance does \emph{not} require decoupling of null-state, but, the theories in which null-states decouple are certainly more solvable or simpler. We know from our experience with $2$-D CFTs that decoupling of Virasoro null-states can even make a theory exactly solvable. So the existence of Virasoro or super-rotation in $D=3+1$ perhaps points to a far more solvable $3+1$ dimensional theory. So it will be interesting to explore the role Virasoro null states in $D=3+1$. This will of course require an understanding of the central term.  

\section{Global conservation laws from boundary conditions}\label{inv}
This is a consistency check for the boundary conditions \eqref{moff} and \eqref{Hf} that we have used for solving the differential equations. One may be skeptical about this this boundary because this was derived under the assumption that inversion is a symmetry. The following demonstration shows its reasonableness. 

\subsection{$U(1)$ charge conservation}
For $U(1)$ we obtained the following boundary condition by using inversion and the fact that $S_a(x)$ is a $(\D=1,l=1)$ primary,
\be\label{bu}
\bra{\{\omega_i,x_i,q_i,out\}} S^{out}_a(x)\ket{0,in} \xrightarrow{\text{$x\rightarrow\infty$}}  \frac{1}{x^2} I_{ab}(x) M_b(\{\omega_\alpha,x_\alpha,q_\alpha\}) + O(\frac{1}{x^3})
\ee
where $M_a$ is a vector without $x$ dependence. This boundary condition can be applied to the Weinberg soft-theorem because the leading soft--photon operator is also a $(\D=1,l=1)$ primary. In terms of the $(\omega,x)$ variables Weinberg's soft photon theorem can be written as \cite{Kapec:2017gsg},
\be\label{wsp}
\bra{\{\omega_i,x_i,q_i,out\}} S^{out}_a(x)\ket{0,in}  \\
= \gamma \bigg( \sum_{i\in out} q_i \frac{2 (x-x_i)_a}{(x-x_i)^2} \bigg)  \bra{\{\omega_i,x_i,q_i,out\}}\ket{0,in}
\ee
where all the particles are assumed to be outgoing. The soft-factor is usually written in terms of the soft-photon polarization vector for which a particular choice has been made in this formula. Charge conservation is then derived from the fact \cite{Weinberg:1964ew} that the Lorentz invariance of the $S$-matrix also requires it to be invariant under on-shell gauge transformation of the soft-photon polarization vector, i.e, $\epsilon_\mu(q) \rightarrow \epsilon_\mu(q) + \alpha q_\mu$, where $q_\mu$ is the $4$-momentum of the soft-photon. Now, for the purpose of deriving global charge conservation, instead of the on-shell gauge-invariance one can use the boundary condition \eqref{bu}. This is also more natural in our context. Of course, on-shell gauge invariance and the falloff condition in \eqref{bu} are both consequences of Lorentz invariance of the $S$-matrix. We would like to note that this is similar to the situation in 2-D CFT where the fall-off condition $T(z)\sim 1/z^4$ gives the Ward identities for the global conformal group starting from the correlation function of $T(z)$ and a set of primaries. 

Let us now find out the large-$x$ behavior of \eqref{wsp}. Since only the soft-factor depends on $x$ we simply expand the soft-factor for large-$x$ at fixed $(\omega_i,x_i)$. The leading term is,
\be\label{lu}
\big(2 \sum_iq_i\big)  \frac{x_a}{x^2} \bra{\{\omega_i,x_i,q_i,out\}}\ket{0,in} \sim O(\frac{1}{x})
\ee 
But, \eqref{bu} tells us that the leading term is of $O(1/x^2)$ because $I_{ab}\sim O(1)$. Therefore the leading $O(1/x)$ term in \eqref{lu} should vanish and this gives the global charge conservation,
\be\boxed{
\sum_i q_i =0}
\ee

\subsection{Energy-momentum conservation}
In this case one has to use the leading soft-graviton theorem. The relevant boundary condition is given by \eqref{Hf}
\be\label{bst}
\bra{\{\omega_i,x_i, out\}} h^{out}_{ab}(x) \ket{0,in} \xrightarrow{\text{$x\rightarrow\infty$}}  \frac{1}{x^2} I_{ac}(x) I_{bd}(x) M_{cd}(\{\omega_\alpha,x_\alpha\}) + O(\frac{1}{x^3})
\ee
where $M_{ab}$ is a traceless symmetric tensor without $x$ dependence and $h_{ab}(x)$ is a $(\D=1,l=2)$ primary which can be taken to be the leading soft-graviton operator. Now Weinberg's soft-graviton theorem in $D$ space-time dimensions can be written as \cite{Kapec:2017gsg}, 
\be
\bra{\{\omega_i,x_i, out\}} h^{out}_{ab}(x) \ket{0,in} = \gamma \frac{2}{n} \sum_i \omega_i \bigg( \delta_{ab} - n \frac{(x-x_i)_a(x-x_i)_b}{(x-x_i)^2} \bigg) \bra{\{\omega_i,x_i, out\}}\ket{0,in}
\ee
where $n=D-2$. In this paper we have worked in $D=6$ but this demonstration is valid for any dimension. The boundary condition does not depend on the dimension because in any dimension the leading soft-graviton operator is a $(\D=1,l=2)$ primary. We now expand the soft-factor for large $x$ but fixed and finite $(\omega_i,x_i)$. We also abbreviate the $S$-matrix element \emph{without} the soft-graviton simply as $S_0$. The results are the following : \\

1) $O(1)$ : The $O(1)$ term in the expansion is given by, 
\be
A_0 = \big(\sum_i \omega_i \big) \bigg( \delta_{ab} - n \frac{x_ax_b}{x^2}\bigg) S_0
\ee
Since there is no $O(1)$ term in the boundary condition \eqref{bst} we immediately get the conservation law,
\be
\boxed{
\sum_i \omega_i =0 }
\ee
\\

2) $O(1/x)$: The order $O(1/x)$ term in the expansion can be written as (up to constant multiplicative factors),
\be
A_1 = \frac{1}{x^2} \sum_i \omega_i \bigg( x_a x_{ib} + x_b x_{ia} - 2 x_a x_b \frac{x_i \cdot x}{x^2} \bigg) S_0  
\ee
Again since there is no $O(1/x)$ term in the boundary condition \eqref{bst} we get,
\be\label{1}
\sum_i \omega_i \bigg( x_a x_{ib} + x_b x_{ia} - 2 x_a x_b \frac{x_i \cdot x}{x^2} \bigg) = 0
\ee
Now let us define,
\be
X_a = \sum_i \omega_i x_{ia}, \quad n_a = \frac{x_a}{|x|}
\ee
In term of $X$ and $n$ we can write,
\be\label{2}
n_a X_b + n_b X_a - 2 n_an_b n\cdot X =0 
\ee
Now the boundary condition was derived under the condition that the point $x$ can be varied arbitrarily while keeping all the $(\omega_i,x_i)$ fixed. Therefore we can take, for example, $n=(1,0,0.....)$ and set $a=1,b=p\neq 1$ in \eqref{2}. This immediately gives $X_p=0, \forall p\neq 1$. Similarly taking $n$ to be $(0,1,0,....)$ and setting $a=2, b=1$ we get $X_1=0$. So we have the conservation laws,
\be
X=0 \rightarrow \boxed{\sum_i \omega_i x_{ia}=0}, \quad a = 1,2,...,n
\ee
\\

3) $O(1/x^2)$ : At this order we get, 
\be
A_2 =  - \frac{n_an_b}{x^2} \sum_i\omega_i x_i^2 S_0 + \frac{1}{x^2} I_{ac}(x)I_{bd}(x) \sum_i \omega_i x_{ic} x_{id} S_0
\ee
Comparing this with the boundary condition \eqref{bst} we get the conservation law,
\be
\boxed{
\sum_i \omega_i x_i^2 =0 }
\ee
Since the null momenta are parametrized as, 
\be
p_i = \omega_i (1 + \vec x_i^2 , 2 \vec x_i , 1- \vec x_i^2)
\ee
The three conservation laws are nothing but the energy-momentum conservation laws. 

\section{Resemblance to string theory : Some speculations}
One of the very basic facts of string theory, which comes out of the quantization of fundamental strings, is the correspondence between null-states \footnote{In earlier sections we have referred to null-states as primary descendants.} and space-time gauge symmetry. Let us briefly describe this in case of photon in the open-string spectrum. Thinking of the open-string as the holomorphic half of the closed string we can write the photon vertex operator as, $V= i e_{\mu}(p) \pa_z X^{\mu} e^{ip.X}$, where $e_{\mu}(p)$ is the photon polarization vector. Here normal-ordering is implied. Now physical state condition gives the two equations,
\be\label{psc}
p^2 =0 , \quad p^{\mu} e_{\mu}(p) =0
\ee 
Now there is a null-state at this level given by $\tilde V = i p_{\mu} \pa_z X^{\mu} e^{ip.X} = L_{-1} e^{ip.X}$. Then the gauge transformation is given by $V \rightarrow V + \alpha \tilde V$ where $\alpha$ is a number. This amounts to the on-shell gauge transformation of the photon polarization vector, $e_\mu(p) \rightarrow e_\mu(p) + \alpha p_{\mu}$. Since $p_\mu$ is a null-vector, the physical sate condition $e.p =0$ is invariant under the transformation $e_\mu(p) \rightarrow e_\mu(p) + \alpha p_{\mu}$. Similar thing happens at higher level also. 

Now this is somewhat similar to what we find this paper. For example, let us consider the leading soft-photon operator $S_a(x)$ in space-time dimension $D=6$. This is a $(\D=1,l=1)$ primary. We have seen that the $U(1)$ global symmetry and (Lorentz) conformal invariance forces the operator $S_a$ to satisfy the equation,
\be
\pa_a(\pa_a S_b - \pa_b S_a) =0
\ee
This can be thought of as the analog of the \emph{physical state condition} \eqref{psc}. Now $\pa_a(\pa_a S_b - \pa_b S_a)$ itself is a primary descendant or null-state, but this is \emph{not} the analog of the null-state $L_{-1}e^{ip.X}$. The analog of $L_{-1}e^{ip.X}$ is the \emph{unique} $(\D=1,l=1)$ null-state $\pa_a S^0$, where $S^0$ is a $(\D=0,l=0)$ primary. Therefore we can speculate on a formal similarity of the following form,
\be
\pa_a(\pa_a S_b - \pa_b S_a) =0   \sim p^2 =0 , \quad p^{\mu} e_{\mu}(p) =0 
\ee
\be
\pa_a S^0(x) \sim L_{-1}e^{ip.X} 
\ee
\be\label{lgt}
S_a \rightarrow \tilde S_a = S_a + \pa_a S^0 \sim e_{\mu}(p) \rightarrow \tilde e_\mu(p) = e_\mu(p) + \alpha p_\mu 
\ee
\be
\pa_a(\pa_a \tilde S_b - \pa_b \tilde S_a) =0 \sim \tilde e_\mu(p) . p=0
\ee
The last equation means that the physical state condition is invariant under gauge transformations. In our case the the transformation \eqref{lgt} is somehow related to the \emph{large} $U(1)$ gauge transformations at infinity. One check of this parameter counting which matches. Proving this will be very interesting. 

The same thing can be said about the leading soft-graviton operator $h_{ab}(x)$ which is a $(\D=1,l=2)$ primary. In this case the "physical-state" condition is given by,
\be\label{psc2}
\pa^2 h_{ab} - \frac{2}{3} \bigg( \pa_a \pa_c h_{cb} + \pa_b \pa_c h_{ca} \bigg) +\frac{1}{3} \delta_{ab} \pa_c\pa_d h_{cd} = 0
\ee
Now there is a \emph{unique} $(\D=1,l=2)$ \emph{primary descendant or null-state}, 
\be
\tilde N_{ab}(x) = (\pa_a \pa_b - \frac{1}{4} \delta_{ab} \pa^2) S^{-1}(x)
\ee
which is a level--$2$ descendant of the $(\D=-1,l=0)$ primary $S^{-1}(x)$. So the "gauge-transformation" should be, 
\be\label{gsut}
h_{ab}(x) \rightarrow \tilde h_{ab}(x) = h_{ab}(x) + \alpha \tilde N_{ab}(x) = h_{ab}(x) + \alpha (\pa_a \pa_b - \frac{1}{4} \delta_{ab} \pa^2) S^{-1}(x)
\ee
The "physical state" condition \eqref{psc2} is invariant under the "gauge transformation" \eqref{gsut}, i.e, 
\be
\pa^2 \tilde h_{ab} - \frac{2}{3} \bigg( \pa_a \pa_c \tilde h_{cb} + \pa_b \pa_c \tilde h_{ca} \bigg) +\frac{1}{3} \delta_{ab} \pa_c\pa_d \tilde h_{cd} = 0
\ee
The same considerations apply to $D=4$ space-time dimensions also. Now, it remains to be seen that to what extent we can take such analogy seriously. We hope to address some of these issues in future. 

Before we conclude, let us make some general comments. The reader may have observed a pattern here. We can start with a $(\D,l)$ primary, $S^{\D}_{a_1a_2...a_l}$ and ask, what kind of (large) gauge transformation is it potentially connected to? If, for the time being, we subscribe to the rule of thumb that \emph{null state corresponds to large gauge transformation}, then this problem can be completely solved using the results of \cite{Penedones:2015aga}. What we have to do is to look for $(\D,l)$ primary descendant or null-state. The null-state should have dimension $\D$ and spin $l$ because we have to add the null-state to $S^{\D}_{a_1a_2...a_l}$. Let us do it for a $(\D=0,l=2)$ primary, which could be a sub-leading soft-graviton As we have already discussed in section \eqref{null}, according to \cite{Penedones:2015aga} there are three types of primary descendants : Type-$1$, type-$2$ and type-$3$. Now one can easily check that a $(\D=0,l=2)$ null-state can only be of type-$1$. Now suppose the $(\D=0,l=1)$ null-state is the level-$n$ descendant of the $(\delta,l')$ primary. Since $(\D=0,l=2)$ is of type-$1$ we have, 
\be
\delta = 1- n - l' , \quad  \delta + n = \D = 0 , \quad  n + l' = 2 
\ee
The unique solution is given by $(\delta = -1, l' =1)$ and $n=1$. So if we denote the $(\delta=-1,l'=1)$ primary by $\xi_a(x)$ then the $(\D=0,l=2)$ null-state is given by,
\be
D_{ab}(x) = \pa_a \xi_b + \pa_b \xi_a - \frac{1}{2} \delta_{ab} \pa\cdot\xi
\ee
therefore the probable "(large) gauge transformation" should be given by,
\be\label{cd}
S^{0}_{ab} \rightarrow S^0_{ab} + D_{ab} = S^0_{ab}(x) + \pa_a \xi_b + \pa_b \xi_a - \frac{1}{2} \delta_{ab} \pa\cdot\xi
\ee
where $S^0_{ab}$ is the sub-leading soft-graviton operator. We have done this for $D=6$ but the same thing applies to $D=4$ with obvious changes. So, we can see that now, the transformation parameter is a vector field $\xi_a(x)$ and the possible gauge transformation \eqref{cd} is the change in the traceless part of a metric tensor under diffeomorphism generated by $\xi_a$. This is also the linearised "gauge transformation" of Weyl gravity expanded around a flat background. Let us now discuss what happens in $D=4$ space-time dimensions. 

In order to see the unique feature of $D=4$ let us extract the global transformations from the (large) gauge transformations. For example, in the case of the $U(1)$, \emph{global $U(1)$} can be obtained by setting the (large) gauge transformation to zero, i.e, $\pa_a \phi =0$. This gives the expected answer, $\phi = const$. For supertranslation this is more non-trivial. In this case the (large) gauge transformation is given by \eqref{p2d}. Setting it to zero we get, 
\be
(\pa_a\pa_b - \frac{1}{2} \delta_{ab} \pa^2) f =0 
\ee
In complex coordinates this gives us two equations,
\be
\pa_z^2 f =0 , \quad \pa_{\bar z}^2 f =0
\ee
whose solutions are given by, 
\be
f(z,\bar z) = A + B z + \bar B \bar z + C z\bar z
\ee
where we have used the fact that $f(z,\bar z)$ is real. So the global transformation has four real parameters. This is the expected answer because the four global space-time translations act on the (retarded) Bondi coordinate $U$ as, 
\be
U \rightarrow U + f(z,\bar z) = U +  A + B z + \bar B \bar z + C z\bar z 
\ee
Our $U$ is related to the usual Bondi retarded coordinate $u$ as, $U = u(1+z\bar z)$. This scaling corresponds to going from the spherical to the planar coordinates. 

Let us now study the case of the sub-leading soft-graviton. In $D=6$ or $n=4$, the potential candidate for the (large) gauge transformation is given by \eqref{cd}. So the global transformations in $D=6$ or $n=4$ are given by, 
\be
\pa_a \xi_b + \pa_b \xi_a - \frac{1}{2} \delta_{ab} \pa\cdot\xi = 0, \quad a,b = 1,2,3,4
\ee
This is the conformal Killing equation on $R^4$. So the global transformations form the conformal group of $R^4$ given by $SO(5,1)$, which is also the Lorentz group in $D=6$. This is consistent with the fact that the Lorentz group acts on the celestial sphere by conformal transformations. So the \emph{sub-leading soft-graviton in $D=6$ is expected to be related to Lorentz transformations}. 

Now we can clearly understand the unique place of $D=4$. In $D=4$ or $n=2$ the global transformations are generated by the vector fields satisfying, 
\be
\pa_a \xi_b + \pa_b \xi_a - \delta_{ab} \pa\cdot\xi = 0 , \quad a,b = 1,2
\ee
As is well-known, the conformal Killing equation on $R^2$ has infinite number of solutions. Therefore, instead of recovering the finite dimensional Lorentz group $SO(3,1)$, we obtain the infinite dimensional Wit or Virasoro algebra. Happily, this is the expected answer \cite{Barnich:2011ct, Kapec:2016jld} ! 

Although we seem to have a purely \emph{algebraic} way of \emph{guessing} the right connection, this procedure does not, so far, shed any light on the physical interpretation. The physical interpretation is indirect. But what it clearly shows is that the conformal symmetry and its representation theory, together with the infinite dimensional global symmetry is more powerful than it seems. We hope to return to the physical aspects of this approach in future.

\section{Acknowledgement}
SB would like to thank the string theory group at TIFR, Mumbai for hospitality when the work was in progress. We would also like to thank Abhijit Gadde, Gautam Manda, Shiraz Minwalla, Sandip Trivedi and all the participants of the Quantum Space-time seminar for helpful questions and discussions. We would also like to thank Alok Laddha for helpful correspondence. SB would like to thank the participants of AdS/CFT at  20 and Beyond for useful discussion on related matters. This research was supported in part by the International Centre for Theoretical Sciences (ICTS) during a visit for participating in the program - AdS/CFT at 20 and Beyond (Code: ICTS/adscft20/05). 

\section{Appendix}
\subsection{U(1)}

	For the  $(\Delta=-3,l=1)$ primary $S^{-3}_a \equiv S_a$, the primary descendant which needs to be set to zero is given by, 
	\begin{eqnarray}
	(T_1)_{abcd}=\frac{1}{180}(144 \partial_{\text{$b $}} \partial_{\text{$c $}} \partial_{\text{$d $}} S_{\text{$a $}}+144 \partial_{\text{$a $}} \partial_{\text{$c$}} \partial_{\text{$d $}} S_{\text{$b $}}+144 \partial_{\text{$a $}} \partial_{\text{$b $}} \partial_{\text{$d $}} S_{\text{$c$}}+144 \partial_{\text{$a $}} \partial_{\text{$b $}} \partial_{\text{$c $}} S_{\text{$d $}}\\-18 \partial^2 \partial_{\text{$d $}} S_{\text{$a $}} \delta_{\text{$b $} \text{$c $}}-18 \partial^2 \partial_{\text{$d$}} S_{\text{$b $}} \delta_{\text{$a $} \text{$c $}}-18 \partial^2 \partial_{\text{$d $}} S_{\text{$c $}} \delta_{\text{$a$} \text{$b $}}-18 \partial^2 \partial_{\text{$b $}} S_{\text{$a $}} \delta_{\text{$c $} \text{$d $}}\nonumber\\-18 \partial^2 \partial_{\text{$a $}} S_{\text{$b $}} \delta_{\text{$c $} \text{$d $}}-18 \partial^2 \partial_{\text{$c $}}
	S_{\text{$a $}} \delta_{\text{$b $} \text{$d $}}-18 \partial^2 \partial_{\text{$c $}} S_{\text{$b $}} \delta_{\text{$a$} \text{$d $}}-18 \partial^2 \partial_{\text{$a $}} S_{\text{$c $}} \delta_{\text{$b $} \text{$d $}}\nonumber\\-18 \partial^2 \partial_{\text{$b $}} S_{\text{$c $}} \delta_{\text{$a $} \text{$d $}}-18 \partial^2 \partial_{\text{$a $}}
	S_{\text{$d $}} \delta_{\text{$b $} \text{$c $}}-18 \partial^2 \partial_{\text{$b $}} S_{\text{$d $}} \delta_{\text{$a$} \text{$c $}}-18 \partial^2 \partial_{\text{$c $}} S_{\text{$d $}} \delta_{\text{$a $} \text{$b $}}\nonumber\\-36 \partial_{\text{$a $}} \partial_{\text{$d $}} \delta_{\text{$b $} \text{$c $}} \left(\partial   S\right)-36
	\partial_{\text{$b $}} \partial_{\text{$d $}} \delta_{\text{$a $} \text{$c $}} \left(\partial   S\right)-36
	\partial_{\text{$c $}} \partial_{\text{$d $}} \delta_{\text{$a $} \text{$b $}} \left(\partial   S\right)\nonumber\\-36 \partial_{\text{$a $}} \partial_{\text{$b $}} \delta_{\text{$c $} \text{$d $}} \left(\partial   S\right)\nonumber-36
	\partial_{\text{$a $}} \partial_{\text{$c $}} \delta_{\text{$b $} \text{$d $}} \left(\partial   S\right)-36
	\partial_{\text{$b $}} \partial_{\text{$c $}} \delta_{\text{$a $} \text{$d $}} \left(\partial   S\right)\nonumber\\+12 \partial^2 \delta_{\text{$a $} \text{$d $}} \delta_{\text{$b $} \text{$c $}} \left(\partial   S\right)+12
	\partial^2 \delta_{\text{$a $} \text{$c $}} \delta_{\text{$b $} \text{$d $}} \left(\partial   S\right)+12
	\partial^2 \delta_{\text{$a $} \text{$b $}} \delta_{\text{$c $} \text{$d $}} \left(\partial   S\right)\nonumber)
	\end{eqnarray}
Here $\pa S = \pa_a S_a$ . 	
	
Now $O(x)$ turns out to be a descendant of $T_1$ given by,
	\begin{equation}
		O(x)=\partial_a \partial_b \partial_c \partial_d (T_1)_{abcd}
	\end{equation}
Therefore setting $T_1$ to zero gives us the trivial solution $O(x)=0$. 

\subsection{Supetranslation}	
	
	Similarly for supertranslation we are interested in $l=2$ operators with $\Delta=-3$ and $-5$.
	
For the $(\Delta=-3,l=2)$ operator $S^{-3}_{ab}\equiv S_{ab}$, the primary descendant which needs to be set to zero is given by, 
	\begin{eqnarray}
	(T_1)_{abcd}=\frac{1}{240}(96 \partial_{\text{$c $}} \partial_{\text{$d $}}S_{\text{$a $}\text{$b $}}+96 \partial_{\text{$b $}} \partial_{\text{$d$}} S_{\text{$a $}\text{$c $}}+96 \partial_{\text{$a $}} \partial_{\text{$d $}} S_{\text{$b $}\text{$c$}}
	\\
	+96 \partial_{\text{$b $}} \partial_{\text{$c $}} S_{\text{$a $}\text{$d $}}+
	\left.96 \partial_{\text{$a $}} \partial_{\text{$c $}} S_{\text{$b $}\text{$d $}}+96 \partial_{\text{$a $}} \partial_{\text{$b$}} S_{\text{$c $}\text{$d $}}\right.
	\nonumber\\
	+8 \delta_{\text{$a $} \text{$d $}} \delta_{\text{$b $} \text{$c $}} \left(\partial \left(\partial   S\right) \right)+8 \delta_{\text{$a $} \text{$c $}} \delta_{\text{$b $} \text{$d $}} \left(
	\partial	\left(\partial   S\right) \right)+\nonumber
	8 \delta_{\text{$a $} \text{$b $}} \delta_{\text{$c $} \text{$d $}} \left( \partial\left(\partial  
	S\right) \right)
	\nonumber\\
	-12 \partial^2 S_{\text{$a $}\text{$b $}} \delta_{\text{$c $} \text{$d $}}-12
	\partial^2 S_{\text{$a $}\text{$c $}} \delta_{\text{$b $} \text{$d $}}-
	12 \partial^2 S_{\text{$b $}\text{$c $}} \delta_{\text{$a $} \text{$d $}}\nonumber\\
	-12 \partial^2 S_{\text{$a $}\text{$d $}} \delta_{\text{$b $} \text{$c $}}-12 \partial^2 S_{\text{$b $}\text{$d $}} \delta_{\text{$a$} \text{$c $}}-12 \partial^2 S_{\text{$c $}\text{$d $}} \delta_{\text{$a $} \text{$b $}}\nonumber
	\\-
	24 \partial_{\text{$b $}}  \delta_{\text{$c $} \text{$d $}} \left(\partial   S\right)_{\text{$a $}}
	-24	\partial_{\text{$a $}}  \delta_{\text{$c $} \text{$d $}} \left(\partial   S\right)_{\text{$b $}}
	-24\partial_{\text{$c $}} \delta_{\text{$b $} \text{$d $}} \left(\partial   S\right) _{\text{$a $}}\nonumber
	\\-
	24 \partial_{\text{$c $}}  \delta_{\text{$a $} \text{$d $}} \left(\partial   S\right)_{\text{$b $}}-24
	\partial_{\text{$a $}}  \delta_{\text{$b $} \text{$d $}} \left(\partial   S\right)_{\text{$c $}}-24
	\partial_{\text{$b $}}  \delta_{\text{$a $} \text{$d $}} \left(\partial   S\right)_{\text{$c $}}\nonumber\\-
	24 \partial_{\text{$d $}}  \delta_{\text{$b $} \text{$c $}} \left(\partial   S\right)_{\text{$a $}}-24
	\partial_{\text{$d $}}  \delta_{\text{$a $} \text{$c $}} \left(\partial   S\right)_{\text{$b $}}-24
	\partial_{\text{$d $}}  \delta_{\text{$a $} \text{$b $}} \left(\partial   S\right)_{\text{$c $}}\nonumber\\-
	24 \partial_{\text{$a $}}  \delta_{\text{$b $} \text{$c $}} \left(\partial   S\right)_{\text{$d $}}-24
	\partial_{\text{$b $}}  \delta_{\text{$a $} \text{$c $}} \left(\partial   S\right)_{\text{$d $}}-24
	\partial_{\text{$c $}}  \delta_{\text{$a $} \text{$b $}} \left(\partial   S\right)_{\text{$d $}}
	)\nonumber
	\end{eqnarray}
where $\pa(\pa S) = \pa_a\pa_b S_{ab}$ and $(\pa S)_a = \pa_b S_{ab}$.
	
Now $O(x)$ is a descendant of $T_1$ given by,
\begin{equation}
O(x)=\partial^2 \partial_{\text{$a $}}\partial_{\text{$b $}}\partial_{\text{$c $}}\partial_{\text{$d $}} (T_1)_{abcd}
\end{equation}
Therefore setting $T_1$ to zero gives the trivial solution $O(x)=0$.


The same consideration applies to the $(\D=-5,l=2)$ primary.

\end{document}